\documentclass[11pt,onecolumn]{article} 
\usepackage{graphicx}
\usepackage{floatrow}

\usepackage{psfig}

\usepackage{deluxetable}
\usepackage{graphicx}	
\usepackage{amsmath}	
\usepackage{amssymb}
\usepackage{times}
\usepackage{latexsym}
\usepackage{fnpos}
\usepackage{scrextend}
\usepackage{scalerel}
\usepackage{hyperref}
\usepackage{tabularx}
\usepackage{xspace}
\usepackage{enumerate}
\usepackage{longtable}
\usepackage{lscape}
\def\aj{Astron. J.}
\def\araa{Ann. Rev. Astron. Astrophys.}
\def\apj{Astrophys. J.}
\def\apjs{Astrophys. J. Suppl.}
\def\aap{Astron. Astrophys.}
\def\aaps{Astron. Astrophys. Suppl.}
\def\mnras{Mon. Not. R. Astron. Soc.}
\def\nat{Nature}
\def\JApA{J. Astron. Astrophys.}
\def\pasp{Publ. Astron. Soc. Pacific} 
\def\pasa{Pub. Astron. Soc. Aus.}
\def\rmxaa{RxMAA} 

\def\JATIS{J.Astron.Telesc. Instrum. Syst.}
\def\procspie{Proc. SPIE}

\def\aaps{A\&AS} 
\def\Msngr{The Messenger}
\def\sci{Science} 

\usepackage{amsmath} 

\usepackage{authblk}

\usepackage{cite}
\usepackage[numbers,super,sort]{natbib}

\usepackage{notocite}

\usepackage{hyperref}

\title{AstroSat detection of Lyman continuum emission from a z=1.42 galaxy}

\author[1]{{Kanak Saha}  \footnote{Email: kanak@iucaa.in}}
\author[1]{Shyam N. Tandon}
\author[2]{Charlotte Simmonds}

\author[2]{Anne Verhamme}
\author[1]{Abhishek Paswan}
\author[2]{Daniel Schaerer}
\author[3]{Michael Rutkowski}
\author[4]{Anshuman Borgohain}
\author[5]{Bruce Elmegreen}
\author[6,7]{Akio K. Inoue}
\author[8]{Francoise Combes}
\author[9]{Debra Elmegreen}
\author[10]{Mieke Paalvast}

\affil[1]{Inter-University Centre for Astronomy and Astrophysics, PostBag 4, Ganeshkhind, Pune-411007, India}
\affil[2]{Observatoire de Genève, Universitéde Genève, 51 Ch. des Maillettes, 1290, Versoix, Switzerland}
\affil[3]{Department of Physics \& Astronomy,  Minnesota State University-Mankato, Trafton Science Center 141, Mankato, MN 56001, USA}
\affil[4]{Department of Physics, Tezpur University, Napaam 784028, India}
\affil[5]{IBM Research Division, T. J. Watson Research Center, 1101 Kitchawan Road, Yorktown Heights, NY 10598, USA}
\affil[6]{Department of Physics, School of Advanced Science and Engineering, Waseda University, 3-4-1 Okubo, Shinjuku, Tokyo 169-8555, Japan}
\affil[7]{Waseda Research Institute for Science and Engineering, Waseda University, 3-4-1, Okubo, Shinjuku, Tokyo 169-8555, Japan}
\affil[8]{Observatoire de Paris, LERMA, College de France, CNRS, PSL Univ., Sorbonne University, UPMC, Paris, France}
\affil[9]{Department of Physics \& Astronomy, Vassar College, Poughkeepsie, NY 12604, USA}
\affil[10]{Leiden Observatory, Leiden University, PO Box 9513, 2300 RA, Leiden, The Netherlands}

\date{}

\begin{document}

\maketitle

{\bf 
One of the outstanding problems of current observational cosmology is to understand the nature of sources that produced the bulk of the ionizing radiation after the Cosmic Dark Age. Direct detection of these reionization sources \cite{Stiavellietal2004} is practically infeasible at high redshift due to the steep decline of intergalactic medium transmission \cite{Madau1995,Inoueetal2014}. However, a number of low-redshift analogs emitting Lyman continuum at~900\AA ~restframe are now detected at $z< 0.4$  [\citen{Leitetetal2013,Borthakuretal2014,Izotovetal2016a,Leithereretal2016,Izotovetal2018}] and there are detections in the range ${\bf 2.5}< z< 3.5$ [\citen{Shapleyetal2016,Vanzellaetal2016,Bianetal2017,Vanzellaetal2018,Steideletal2018,Fletcheretal2019}] also.
Here, we report the detection of Lyman continuum emission with a high escape fraction (>20\%) from a low-mass clumpy galaxy at z=1.42, in the middle of the redshift range where no detection has been made before and near the peak of the Cosmic Star-formation history\cite{MadauDickinson2014}. The observation was made in the Hubble Extreme Deep field\cite{Illingworthetal2013} by the wide-field Ultra-Violet-Imaging Telescope\citep{Tandonetal2017a} on-board AstroSat\cite{Singhetal2014}. This is the first detection of Extreme Ultraviolet radiation from a distant galaxy at a rest-frame wavelength of 600Å, and it opens up a new window to constrain the shape of the ionization spectrum. Further observations with AstroSat should significantly increase the sample of Lyman continuum leaking galaxies at Cosmic Noon.}

\bigskip

\begin{figure*}
\begin{flushleft}
\rotatebox{0}{\includegraphics[width=1.\textwidth]{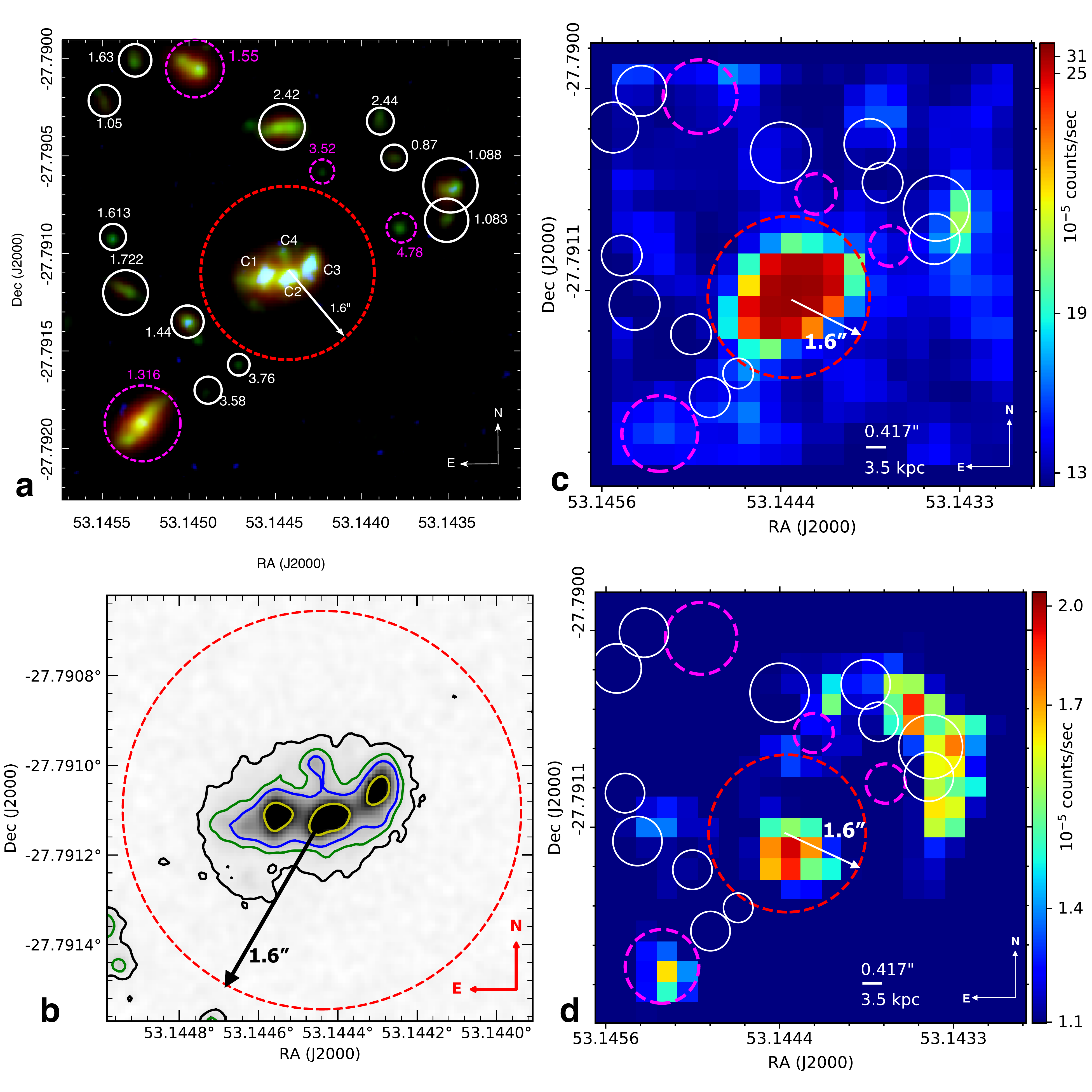}}
\vspace{-1.2cm}
\caption{{\bf Detection of the source in the AstroSat.}
\noindent ({\bf a}): HST colour composite (F275+F606W+F160W) of AUDFs01 in the Hubble Xtreme Deep Field. All 14 objects around the galaxy are marked by coloured circles - magenta (spec-z) and white (photo-z). ({\bf b}): Contour image of the galaxy in F606W band. The outermost contour is at $3\sigma~(27.25 $~mag~arcsec$^{-2}$) level, the next level is drawn at $12\sigma$ level - encompassing all the clumps, C1, C2, C3 and C4. The contours in blue and yellow are at $24$ and $96\sigma$ level. AstroSat observation of AUDFs01 in the NUV ({\bf c}) and FUV ({\bf d}) band, both images are smoothed with a Gaussian kernel of radius=3~pixels. The brightest pixel within 1.6" of the FUV image has $4.43\times10^{-5}$~ct/s. The arc-like feature in the FUV image is due to the blending of several sources, some with low-redshift, corresponding to non-ionising rest-frame FUV emission. On the bottom left corner, the S/N of the FUV emission from the z=1.316 galaxy is too low to claim a positive LyC detection.}
\label{fig:fig1}
\end{flushleft}
\end{figure*}

Low-mass, compact, actively star-forming emission-line galaxies with high equivalent widths are thought to be promising candidates for sources of escaping Lyman Continuum ($\lambda < 912$\AA, hereafter LyC) photons \cite{Izotovetal2016a,Naiduetal2018}. The galaxy AUDFs01 (RA: 53.1444, Dec: -27.7911) at $z=1.42$ has been selected from the Hubble Extreme Deep field\cite{Illingworthetal2013} having one of the highest H$\alpha$~$\lambda 6563$~\AA,  and [O {\sc iii}]~$\lambda 5007$~\AA ~fluxes, measured by the HST Grism G141 under the 3DHST program \cite{Momchevaetal2016}. Both emission lines have high equivalent widths, EW(H$\alpha$)= $1210$ \AA, EW([O~{\sc iii}]) = $1517$ \AA, indicating the production of a copious amount of ionizing photons in the galaxy with relatively few old stars. 
Here, we present broad-band imaging observations of this galaxy in far-ultraviolet (FUV, $1300 - 1750$ \AA, F154W) and near-ultraviolet (NUV, $2000 - 2800$ \AA, N242W) filters using the UltraViolet Imaging Telescope (hereafter, UVIT \cite{Tandonetal2017a}) onboard AstroSat (see Methods, for details about the AstroSat observation, GT05-240, PI: Kanak Saha). In NUV, the galaxy has a magnitude of $25.6\pm0.1$~AB mag with $S/N=10.03$. In FUV band, the source is detected at the same location as in NUV but with a magnitude of $25.84\pm0.34$~AB~mag and S/N=3.2. In either case, the magnitudes are aperture and foreground dust corrected. The detected source has a background-subtracted flux $6.7 \sigma$ above the local background in the FUV band (see Methods). 
The detection of the galaxy in AstroSat~FUV and NUV along with HST UV/Optical/IR is presented in Fig.~\ref{fig:fig1}. The Lyman break of the galaxy AUDFs01 is redshifted to an observed wavelength $\lambda_{lim}=2188.8$~\AA\ (LyC limit for this galaxy) and the FUV filter probes the galaxy in the rest-frame wavelength range $537 - 723$\AA. This is the first detection of Extreme Ultra-Violet (EUV, $\lambda \sim 600$ \AA\ rest-frame) photons from a distant galaxy. The detection of this source, near the peak of the cosmic star-formation history \cite{MadauDickinson2014}, by AstroSat opens up a window to probing star-forming galaxies with EUV photons. This range of wavelength is where models are least constrained, so our observation puts new constraints on the shape of the ionizing spectrum of stellar populations.



This is the first LyC leaking galaxy in the redshift range ${0.4 < z <  2.5}$ where no Lyman Continuum escape has been detected before. This also happens to be the first distant leaking galaxy with a clumpy morphology.
Note that the ionizing radiation ($\lambda \le 912$~\AA)~from sources at $z < 2.7$ would fall at an observed wavelength $\lambda < 3374$~\AA ~and would be blocked by the upper atmosphere for ground-based observation. Although HST/COS spectrograph could have detected sources at this redshift range, the redshift gap (${ 0.4 < z < 2.5}$) remained barren until now. The far-ultraviolet imaging by AstroSat with a wide field of view may significantly increase the number of LyC emitting galaxies in this previously undetected redshift range as the FUV band can exclusively probe LyC emission from galaxies with $z > 0.97$.

We have identified all the objects around AUDFs01 observed by HST and marked their redshifts from the publicly available 3DHST survey \cite{Momchevaetal2016} and MUSE deep field survey \cite{Baconetal2017}. We found no other object detected within a circle of diameter $3.2"$ centered on AUDFs01; all three objects (two with specz and one with photoz) detected immediately outside this circle are at higher redshift than this galaxy. There is no obvious source of contamination to explain the FUV detection other than coming from the clumpy galaxy AUDFs01. We discuss possible contamination further in the method section.
 
\begin{figure*}
\begin{flushleft}
\rotatebox{0}{\includegraphics[width=1.0\textwidth]{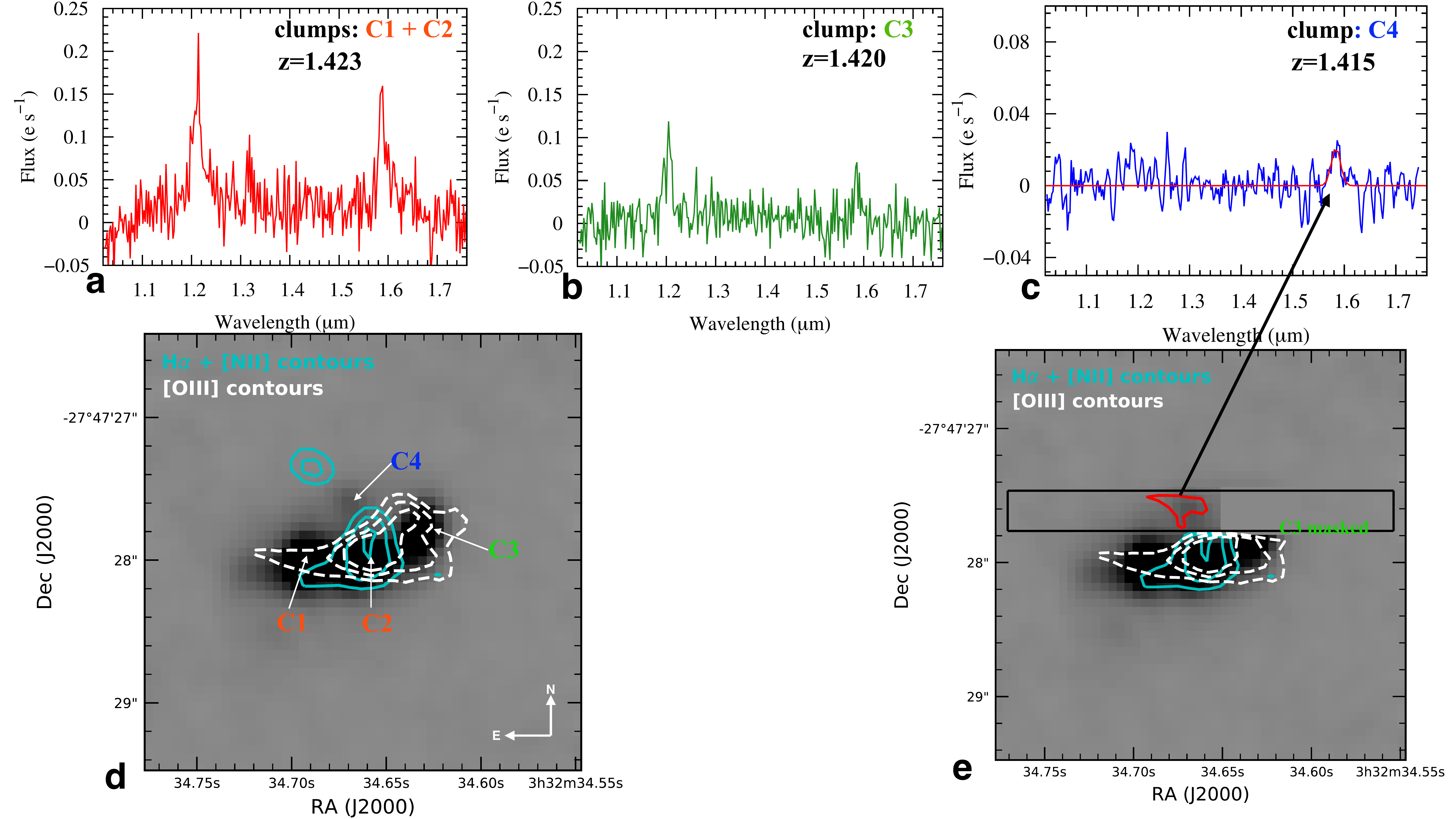}}
\caption{{\bf Emission line mapping and clumps.} ({\bf a}) - 1D spectra extracted from C1+C2. ({\bf b}): for C3 and ({\bf c}): for C4 clump regions (see Fig.~\ref{fig:EDFig1}). Their redshifts are determined by fitting H$\alpha$ line. ({\bf d}): H$\alpha$ and [O~{\sc iii}] contours are overplotted on F160W image at 10$\sigma$ (inner most), 7$\sigma$, and 5$\sigma$ (outer) levels. ({\bf e}): shows the C4 region (red) from which 1D spectrum is extracted; the detected H$\alpha$ line flux is $> 2 \sigma$. The source above clump C1 is detected in H${\alpha}$ at $z=1.433$ but not detected in any other wavelength probed here.}
\label{fig:fig2}
\end{flushleft}
\end{figure*}

\begin{figure*} 
\begin{flushleft}
\rotatebox{0}{\includegraphics[width=1.0\textwidth]{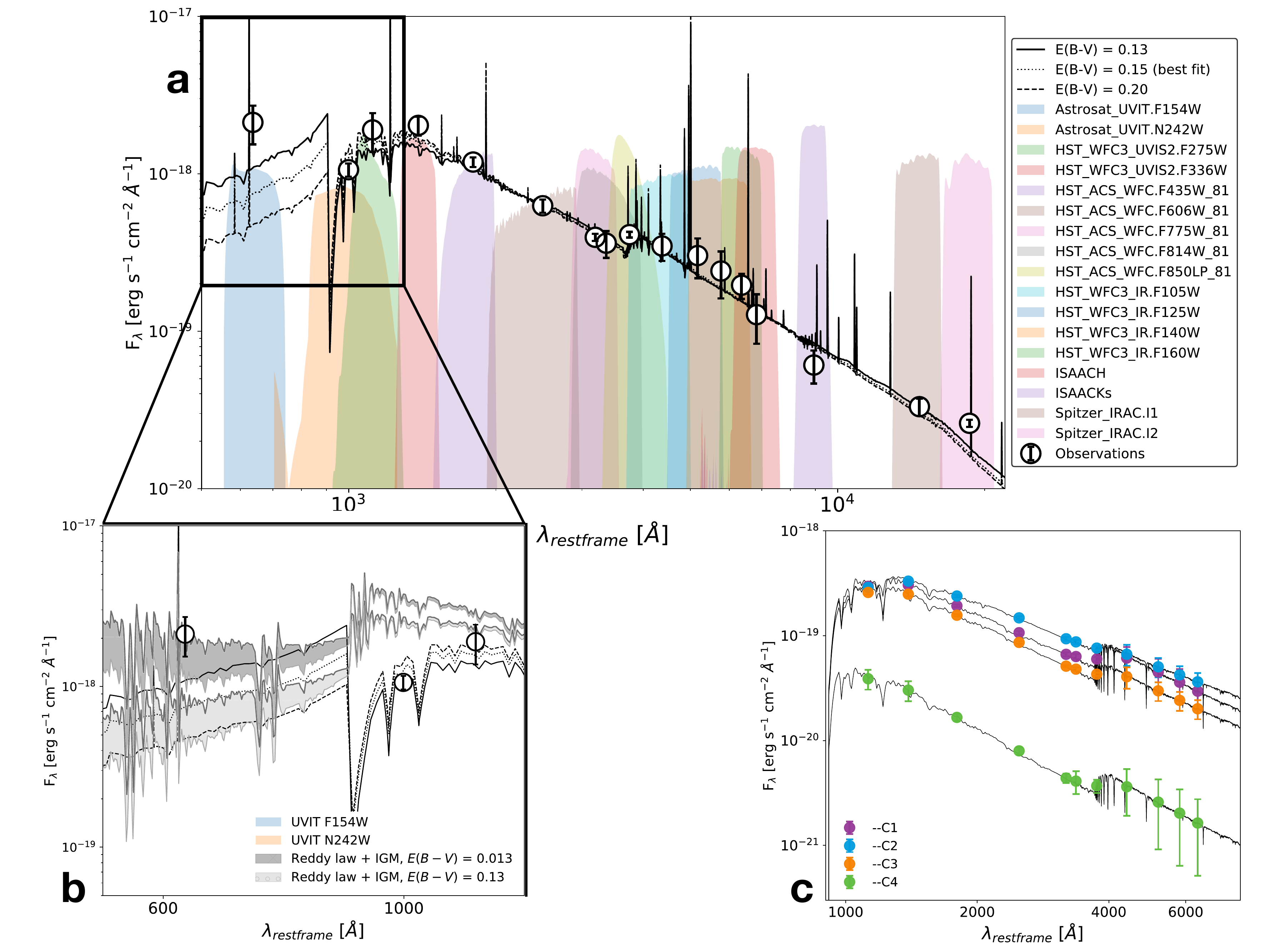}}
\vspace{-0.5cm}
\caption{{\bf Modelling of Spectral Energy Distribution.} ({\bf a}): Stellar + nebular emission models based on BC03 stellar populations are fitted to the observed SED using PCIGALE with varying parameters (see Table~\ref{tab:tab2}). The error bars are denote $1\sigma$ uncertainties (see Table.~\ref{tab:tab3}). The colored shaded regions are the bandpasses of various filters (see side panel) used in this analysis. The best-fit model (dotted line) is one with E(B-V)=0.15 and Z=0.004.  ({\bf b}): the BPASS modelling of the AstroSat UV fluxes with Reddy extinction law. The grey shaded region reflects the IGM distribution with transmission from 0.4 to 1. None of these models alone can reproduce the photometry of AUDFs01 perfectly from rest-frame 600 to 18595\AA. ({\bf c}): Clump SED modelling using EAZY\cite{Brammeretal2008}, primarily to determine their photometric redshift. Plotted lines are the best-fit SED models.}
\label{fig:fig3}
\end{flushleft}
\end{figure*}

As marked in Fig.~\ref{fig:fig1}, the galaxy has 3 bright clumps (C1, C2, C3) and a fainter one (C4) - all appear to be connected by $12\sigma$ surface brightness contour. They are connected in redshift too. In Fig.~\ref{fig:EDFig1}, we show the HST grism G141 image of AUDFs01 and its full spectrum. The redshift of the galaxy is derived using the H$\alpha$ line. To derive the redshift of each clump and the spatial origin of the emission lines, we have constructed an emission-line mapping (see Methods) of the galaxy (Fig.~\ref{fig:fig2}). With that, we have extracted spectra for region~1 (covers C1+C2), region~2 (C3) and region~3 (C4), marked in Fig.~\ref{fig:EDFig1}, and estimate the redshift of each clump using the H$\alpha$ line (Fig.~\ref{fig:fig2}). The H$\alpha$ emission ($> 5 \sigma$) from C1, C2 and C3 are consistent with redshift $z=1.42$. Only when the clumps (C1, C2, and C3) were masked, we could extract H$\alpha$ emission from the clump C4. The fitting of the extracted spectrum (Fig.~\ref{fig:fig2}{\bf{c}}) shows that C4 is at $z=1.415$. Besides, we have also derived the photometric redshift of the clumps by modeling their multi-wavelength Spectral Energy Distribution (SED) constructed from the HST observations from F275W to F160W (a total of 11 passbands for each clump) using EAZY \cite{Brammeretal2008}. The redshift of each clump is found to be $z \simeq 1.42$, consistent with the location of the Balmer break as well as our estimate from the H$\alpha$ emission (see Fig.~\ref{fig:fig3}) - establishing the fact that {\it all four clumps are at the same redshift and integral parts of AUDFs01}.

 Following the N2 method\cite{PettiniPagel2004}, we estimate the Oxygen abundance in the galaxy to be $12 + \log[O/H]=7.99$ - indicating a metal-poor galaxy with $Z \simeq 0.004$. 
From the emission line measurements, we obtain $O{32}$=[O~{\sc iii}] 5007 /[O~{\sc ii}] 3727 = 9.63 and $R{23}$={([O~{\sc iii}]~$\lambda$~5007,4959+ [O~{\sc ii}]~3727])}/H{$\beta$}=10.05, where [O~{\sc ii}]~3727 is the sum of [O~{\sc ii}] doublet resolved by MUSE\cite{Inamietal2017} - these values are comparable to those found in the local LyC leakers \cite{Cardamoneetal2009,Izotovetal2016a}.

The high equivalent widths of the emission lines indicate that AUDFs01 is composed of mostly a young population with an ample abundance of hot O-type stars producing energetic ionizing photons. In fact, with EW(H$\beta$)=$128$ \AA, EW([H$\beta$])+EW([O~{\sc iii}])=$1645$\AA\ ($680$\AA\ rest-frame) is comparable to some of the high-z LyC sources \cite{deBarrosetal2016,Vanzellaetal2018}. Using the H$\alpha$ line flux, we obtain the star formation rate
SFR~$\sim 23 - 40$~M$_{\odot}$~yr$^{-1}$ for the full galaxy, depending on the adopted attenuation correction and neglecting LyC photon escape\cite{Kennicutt1998}.

We estimate an age of $\sim 4 - 6$~Myr for the H~{\sc ii} regions in the galaxy using the H$\beta$ and H$\alpha$ equivalent widths and the Starburst99\cite{Leithereretal1999} model. 
Based on the [N~{\sc ii}] BPT\cite{BPT1981} diagram (Fig.~\ref{fig:EDFig2}) and 7 Ms Chandra X-ray observations\cite{Luoetal2017}, it is inferred that the galaxy does not host an active galactic nucleus (AGN) (see Methods). 

The stellar mass and stellar population age of the galaxy are derived by modeling the multi-wavelength broad-band SED from FUV-to-IR ($1300 \AA - 45000$ \AA) for the whole galaxy (Fig.~\ref{fig:fig3}) using both the PCIGALE \cite{Boquienetal2019} and the BPASS models \cite{Eldridgeetal2017}. The best-fit PCIGALE  model (Fig.~\ref{fig:fig3}) with metallicity $Z=0.004$ yields a continuum colour excess E(B-V)=0.15, and has a total stellar mass of $1.45 \times 10^{9}$~M$_{\odot}$ including a young stellar component of $1.5 \times 10^{8}$~M$_{\odot}$. 
The age of the main stellar population is $\sim 100$~Myr, while the late-burst age is $2$~Myr. The integrated SFR is  $\sim 55$~M$_{\odot}$~yr$^{-1}$. To better model the AstroSat FUV flux, we have run a series of BPASS models (see Methods) with varying column density, opacity of the intergalactic medium (IGM) along with the photoionization code Cloudy \cite{Ferlandetal2013}. We found that a low column density $N_{H} = 10^{17} cm^{-2}$ and younger stellar population of $4 -6$~Myr are preferred to explain the rest-frame FUV flux of this source.
The best-fit clump SED models (Fig.~\ref{fig:fig3}) show that the clump masses are similar to each other e.g., C1: $1.66 \times 10^{8} M_{\odot}$; C2: $2.95 \times 10^{8} M_{\odot}$; C3: $2.31 \times 10^{8} M_{\odot}$, except for C4: $7.30 \times 10^{6} M_{\odot}$. The clump masses of C1, C2 and C3 are relatively higher than those found in local galaxies. However, the clump-to-galaxy mass ratio for AUDFs01 is $\sim 0.47$ - consistent with known clump-to-galaxy mass ratios at this redshift in the GOODS region \cite{Elmegreenetal2009}. 

The low-mass and clumpy morphology of the galaxy favour the escape of the LyC radiation from the galaxy as they offer more surface area for leaking. However, observing such LyC radiation depends on the properties of ISM in the galaxy as well as the IGM between us and the galaxy. When LyC radiation escapes a galaxy, it is redshifted on its way to the observer, as long as it is not absorbed by an intervening HI cloud. The probability of encountering such a cloud depends on the redshift of the emitting source. The IGM absorption is higher at higher redshift ($z> 3.$) while at lower redshift ($z < 0.4$), the transmission is always high. At intermediate redshift, the IGM transmission distribution shows characteristic bi-modality; at the redshift of our object ($z=1.42$), the median IGM transmission is 0.74 and the median optical depth $\tau_{IGM} =0.29$~[\citen{InoueIwata2008,Inoueetal2014}] in the AstroSat F154W band, which probes solely LyC radiation. We consider the IGM bi-modality in our calculation of escape fraction of LyC photons, denoted as $f_{\rm esc}$. 
\par
We use the observed H$\alpha$ flux corrected for attenuation from UV slope (see Table.~\ref{tab:tab1} in Methods) to estimate the number of LyC photons that are absorbed in the galaxy to be $N_{LyC}^{non-esc} = (2.12 \pm 0.22) \times 10^{54}$~s$^{-1}$. The rest-frame luminosity of the galaxy in F154W is $1.68\pm 0.45 \times 10^{43}$~erg~s$^{-1}$ corresponding to escaping LyC photon rate as $N_{LyC}^{emit}={0.54 \pm 0.14} \times 10^{54}$~s$^{-1}$ at the mean wavelength $\sim 600$~\AA. This results in an absolute escape fraction of $f_{\rm esc} =0.2$ for E(B-V)=0.13 and $\tau_{IGM}=0.0$ (see Eq.\ref{eq:fesc2}). 
We obtain a relatively higher escape fraction $f_{\rm esc}=0.5$ from the PCIGALE best-fit SED model which incorporates CLOUDY modeling and where the free parameter $f_{\rm esc}$ was varied from $0.0-0.8$ along with a set of other parameters, see Table~\ref{tab:tab2}. 

We also derive the escape fraction by comparing the observed 
LyC flux to the one expected from stellar population models with the age of $4.5$~Myr estimated from EW(H$\beta$) and EW(H$\alpha$) using the frequently-used relation\cite{Leitetetal2013,Izotovetal2016a,Leithereretal1999}. Considering E(B-V)=0.13 and assuming a transparent IGM, we obtain $f_{\rm esc} \sim 0.31$; while for the median IGM transmission, $f_{\rm esc} \sim 0.42$. Our estimates of $f_{\rm esc}$ are broadly consistent with each other. Based on our calculations, the $z=1.4$ clumpy galaxy is emitting at least $\sim 20$\% of ionizing photons towards the IGM. 

\begin{figure*}[!h]
\begin{flushleft}
\rotatebox{0}{\includegraphics[clip,width=1.0\textwidth]{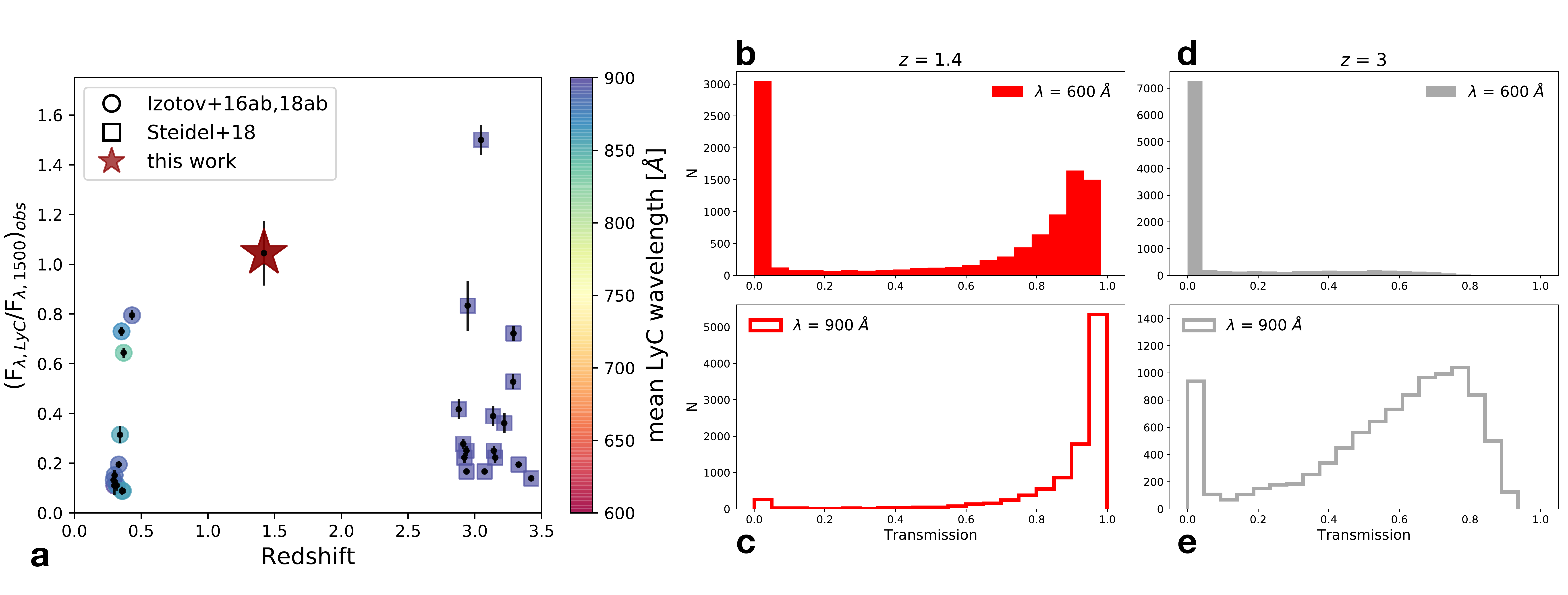}}
\vspace{-1.2cm}
\caption{{\bf Comparison of AUDFs01 with other confirmed LyC detection and IGM distribution.} {\bf a:} Objects with spectroscopically confirmed LyC detection (to-date) are compared with our source found in the redshift desert. The FUV flux detected by AstroSat-UVIT for AUDFs01 at $z\sim 1.42$ corresponds to high energy EUV photons at rest-frame wavelength of $537 - 723$ \AA. The ionizing spectrum of a star-forming galaxy is revealed for the first time over this wavelength window. The error bars represent $1\sigma$ uncertainties on the flux measurements. 
{Distribution of IGM transmissions along 10,000 lines of sights computed using Monte Carlo simulations\cite{Inoueetal2014}, for 600\AA ({\bf b}) and 900\AA ({\bf c}), and for $z\sim1.4$ ({\bf b, c}) and $z\sim3$ ({\bf d, e}).}}
\label{fig:fig4}
\end{flushleft}
\end{figure*}

In Fig.~\ref{fig:fig4}(a), we compile 
 known sources that are spectroscopically confirmed LyC Emitters. In the relatively low-redshift universe ($z < 0.4$), 11 LyC Emitters have been identified among the population of  Green Pea galaxies (GPs), using HST-COS \citep{Izotovetal2016a, Izotovetal2016b, Izotovetal2018, Izotovetal2018b}; and in the more distant universe ($z > 3.0$), 15 individual detections were reported among the population of Lyman Break Galaxies (LBGs), using KECK-LRIS \cite{Steideletal2018}. All flux ratios are given in $F_{\lambda}$. A few more LyC Emitters were reported recently in the literature, such as the strongly lensed Sunburst Arc at $z\sim 2.4$~[\citen{Rivera-Thorsenetal2019}], or Ion3 and Ion4 at $z\sim3$ and $z\sim4$~[\citen{Vanzellaetal2018}], with comparable ${F_{LyC}/F_{1500}}$. 
Our source, observed in FUV and NUV with AstroSat/UVIT \cite{Singhetal2014,Tandonetal2017b}, is unique from two points of view: first, the Extreme UV range of the spectrum of a star-forming galaxy, around $600$\AA\ rest-frame and second, AstroSat is opening up a new redshift range for searches of LyC Emitters at $z\sim 1 - 2$, where the IGM is still fairly transparent (see Fig.~\ref{fig:fig4}(b)). At the redshift of our source, more than 80\% of the lines of sight have a transmission above 80\% at $900$\AA\ rest-frame; and at $600$\AA\ rest-frame, almost 50\% of the lines of sight have a transmission better than 80\%, illustrating the potential of AstroSat to detect efficiently a great number of LyC Emitters over $z\sim1-2$ and so far unexplored rest-frame LyC wavelength ranges. 
Detecting LyC emitters at lower z is important because the fainter sources can be seen at closer distances and there are, in general, many more smaller galaxies than the larger ones. Perhaps, the smaller ones dominated the reionization. Future samples of LyC Emitters, that are likely to be discovered in AstroSat UV Deep field (AUDF) in FUV and NUV bands, have the potential to unveil the nature of the sources responsible for the production of this extreme UV radiation and improve our understanding of the sources that have led to Cosmic Reionization. 



\noindent {\bf Acknowledgement}\\
The deep field imaging data in far and near ultraviolet wavelengths are based on a proposed observation carried out by the AstroSat/UVIT which was launched by the Indian Space Research Organization (ISRO). We thank ISRO for providing such observing facilities. KS and FC acknowledge the support of CEFIPRA-IFCPAR grant through the project no. 5804-1. KS thanks David Sobral for kindly providing the code to make in Fig. 2(d).

\medskip{}
\noindent {\bf Corresponding author}\\
\noindent Correspondence to Kanak Saha (Email: kanak@iucaa.in)\\
\medskip{}

\noindent {\bf Ethics Declarations}\\
\noindent {\large Competing interests}\\

\noindent The Authors declare no competing interests.\\

\newpage

\section{Methods}

A flat $\Lambda$~CDM cosmology with $H_{0}$ = 70 km$s^{-1}$ Mpc$^{-1}$ , $\Omega_{m}$ = 0.3, and $\Omega_{\Lambda}$ = 0.7 was adopted throughout the article. All magnitudes quoted in the paper are in the AB system. 

\subsection{AstroSat observation and other archival data}
The clumpy galaxy AUDFs01 was selected from the Hubble eXtreme Deep Field (XDF), having one of the deepest WFC3/IR images. We first searched the 3DHST \cite{Momchevaetal2016} GOODS-South catalog of 1517 galaxies having spectroscopic redshift and non-zero, positive H$\alpha$, [O~{\sc iii}] and H$\beta$ fluxes. The simultaneous presence of these emission lines is possible only for galaxies in the redshift range $1.1 < z < 1.55$, the so-called redshift desert \cite{Steideletal2004, Renzini2009}. This reduces our sample of 1517 to 948 galaxies. Of these, only 38 galaxies are in the Hubble XDF (denoted as stars symbol in Fig.~\ref{fig:EDFig2}). The clumpy galaxy AUDFs01 is chosen as one having the highest [O~{\sc iii}] and H$\alpha$ fluxes and stellar mass around $\sim 10^{9} M_{\odot}$ - indicating a vigorous star-formation activity.
The photometry for the full galaxy uses archival HST imaging data in the UV(F275W, F336W), Optical(WFC/ACS: F435W, F606W,F775W, F814W and F850LP), and IR (WFC3: F105W, F125W, F140W and F160W). Also, we use the archival VLT/ISAAC observation in the H and Ks-band and Spitzer/IRAC data in the $3.6$ and $4.5$~{$\mu m$}. Fig.~\ref{fig:EDFig3} shows their postage stamp images from GALEX FUV to Spitzer 4.5~$\mu$m band. 

The FUV and NUV imaging data come from AstroSat observations of the GOODS-South field (GT05-240, PI: Kanak Saha). These observations were carried out by UVIT (having a field of view (FOV) of 28' diameter) on-board AstroSat, which performed simultaneous observation in F154W and N242W band during Sept. - Oct. 2016. The observation was carried out for 100 kilosec - which corresponds to $\sim 100$ orbits. During each orbit, FUV and NUV observation in photon counting mode were taken every 33 millisecond - resulting in about 45000 - 50000 frames (in each band) accumulation in a good orbit. The orbit-wise dataset was processed using the official L2 pipeline in which we removed frames that were affected by the cosmic-ray shower and those frames were excluded in the final science-ready images and the subsequent calculation of photometry. This has resulted in $\sim 15\%$ loss of data to science-ready images. In addition to this, there was data loss due to the mismatch of time-stamp on VIS (visual) filter and NUV or FUV filters. The final science-ready image of the AstroSat UV deep Field (AUDF) had a total exposure time of $t_{exp}=63938.5$~sec in FUV and $62341.1$~sec in NUV corresponding to $\sim 63$~AstroSat orbits.

We perform differential astrometry in two steps: first, we use GALEX deep field data of the GOODS-South region. Followed by this step, we use the HST F606W optical image of the GOODS-South as a reference image and redo the astrometric correction. We use an IDL program which takes an input set of matched xpixel/ypixel (from {UVIT} image) and RA/Dec (from F606W image) and perform a TANGENT-Plane astrometric plate solution similar to ccmap task of IRAF. The astrometric accuracy in NUV is found to be $\sim 0.2"$ while for FUV, the RMS$\sim 0.3"$ (note, a pixel $\sim 0.4"$). The photometric calibration is performed with a white dwarf star Hz4; the photometric zero-points\cite{Tandonetal2017a} are 17.78 and 19.81 for F154W and N242W respectively. Once photometric calibration and astrometric correction are successfully applied, we extract an image of the same size as the Hubble XDF from the UVIT full FOV and run SExtractor\cite{BertinArnouts1996} on it. We have extracted all sources that have $S/N \ge 3$ using the following relation:

\begin{equation}
    S/N= \frac{S \times t_{exp}}{\sqrt{S\times t_{exp} + B\times N_{pix} \times t_{exp}}},
 \label{eq:SN}
\end{equation}

In the above equation, S denotes the background-subtracted source flux within a given aperture (having $N_{pix}$ as the number of pixels), B is the background within the same aperture. $t_{exp}$ denotes the exposure time in the same pass-band. In writing the above equation, we have neglected read noise which is null in the CMOS detector used in UVIT. The dark current is about 10 e-/s over the full circular FOV (diameter 28') corresponding to $\sim 7.8\times 10^{-7}$~ct~s$^{-1}$~pix$^{-1}$. Note that the dark current in our detector is about 10 times smaller than the background as estimated below. The background (B) measured directly from the science-ready image includes this dark current contribution.
Once all the sources are detected in the original science image, we mask sources with $S/N \ge 5~\sigma$. Then we re-run SExtractor and remove all the sources $\ge 3~\sigma$ and produce a masked image again. We use this masked image to create a histogram out of N($\simeq$~few 1000) random apertures of 1.6" radius placed all over the image, avoiding the source locations. This histogram appears slightly positively skewed. Furthermore, we remove sources $>2.75~\sigma$ and then carried out the same exercise as above. The histogram on this residual image is found to be nearly symmetric, and fitted with a Gaussian distribution. We reach a global sky surface brightness (mean value) of $28.6$~mag~arcsec$^{-2}$ and $27.8$~mag~arcsec$^{-2}$ in the FUV and NUV-band respectively. {The $3\sigma$ detection limit for point source within 1" radius in FUV is 27.9 AB mag.}
The same procedure has been applied to a smaller cutout (100" x 100") around AUDFs01 to estimate the local background appropriate for the galaxy. The local mean from the Gaussian fit is $B=5.7 \times 10^{-6}$~ct~s$^{-1}$~pix$^{-1}$ (or $28.9$~mag~arcsec$^{-2}$) and a background RMS, $\sigma_{bkg}=6.8\times10^{-6}$~ct~s$^{-1}$~pix$^{-1}$. 

\par
We estimate the source+background flux ($F_{SB}$) within an aperture of radius 1.6" ($N_{pix}=46$) for AUDFs01. By considering the astrometric error of $\sim 0.3"$, we randomly place $25$ such 1.6" apertures about the location of AUDFs01. The mean source+background flux is $<F_{SB}> = (5.71 \pm 0.058)\times 10^{-4}$~ct~s$^{-1}$ (the median $=5.74\times 10.^{-4}$). From this, we obtain the mean background-subtracted source flux as $S =3.1\times 10^{-4}$~ct~s$^{-1}$ corresponding to $26.5$~AB mag without any aperture correction and foreground dust correction in the FUV band. Throughout this paper we have used directly measured fluxes and noise from a given aperture to estimate the S/N. The aperture plus dust corrected magnitudes for AUDFs01 in different bands are presented in Table~\ref{tab:tab3}. With the estimated local background and using eq.~\ref{eq:SN}, we obtain $S/N=3.2$ for AUDFs01 in FUV. The detected source has a flux $6.7 \sigma$ (=S/${\sqrt{N_{pix}}}{\sigma_{bkg}}$) above the local background in the same F154W band.

\subsubsection{Other sources}
 We have examined all the 38 sources in the HST F606W band as mentioned above. We identify only those objects that are clean such that there are no other sources (in particular, ones with $z < 0.97$) present within 1.6" aperture. We found 9 sources that satisfy this criterion and of these, two pointing refer to the same source. So we are left with only 8 sources. Of these 8 sources, only 2 sources have $S/N \ge 3$ detection in the F154W band of AstroSat. Of these, we picked AUDFs01 having the highest {[O~\sc{iii}]} and H$\alpha$ fluxes as well as equivalent widths. The other source at z=1.29 (53.176,-27.773) had lower {[O~\sc{iii}]} flux as well as equivalent width (EW~{[O~\sc{iii}]}=36.9\AA). This source, together with all other upper limits, will be described in a forthcoming paper.  
Note that the galaxy at z=1.316 as shown in Fig.~\ref{fig:fig1}(a) has F154W flux detected at $S/N=2.5$. When we inspected the F606W image carefully, we found another object within a circle of 1.6" of this galaxy; although the object has a photoz z=3.58, we restrict ourselves to spectroscopic redshift only for the purpose of clean LyC detection. None of the sources on the North-West corner (forming an arc-like feature) of F154W image (Fig.~\ref{fig:fig1}) has spectroscopic redshift; all of them have photoz with $z> 0.97$, except one at z=0.87. The FUV emission (e.g., centered on sources at z=1.083 and 1.088) is indeed stronger ($\sim 3~\sigma$) but all these sources are blended in F154W band due to the large FUV PSF, making an arc-like appearance on the smoothed FUV image. The non-leaking source at z=0.87 is a foreground contamination to the rest.

 \subsection{Contamination hypothesis}
{\it An intervening star}:

Could the AstroSat FUV flux be due to a 'contamination' from an intervening star, invisible at longer wavelength? This is extremely unlikely, as it would correspond to the chance alignment of an early type star at Megaparsec distances. Indeed, to explain the observed flux ${F_{1500}\sim 2\times 10^{-18}}$~erg~s$^{-1}$~cm$^{2}$~\AA$^{-1}$ with an unreddened O4V star (with $T_{\rm eff} = 43000$ K, $\log(L/L_{\odot})=5.7$)\cite{Martinsetal2005} - would imply a distance of the order of $\sim 50$~Mpc. Even for much fainter hot stars with similar temperatures, this would imply unrealistically large distances. If the star was significantly
reddened, it would be detected at longer wavelengths and hence be distinguishable as such from the SED and the available spectra.

 {\it An interloper galaxy}:

The possibility of the AstroSat FUV flux to be due a foreground galaxy, invisible at longer wavelength is also unlikely. Since such a contaminating galaxy would be bright in FUV, it would be a star-forming galaxy, with nebular recombination lines emitted from H~{\sc ii} regions.  The Lyman-alpha emission from a foreground Lyman-alpha emitter at 0.069 < z < 0.439 would fall in the AstroSat FUV broad band. Assuming that part of the flux detected by AstroSat would be due to this emission line, we argue that the associated optical nebular lines, H$\alpha$ and the strong [O~{\sc iii}] doublet ($\lambda 4959,\lambda 5007$\AA), would be detected in the MUSE data\cite{Inamietal2017}. We extract a MUSE spectrum within a round 1.6" aperture, to compare with the AstroSat detection, and examined the spectrum in detail. None of the two lines mentioned above are detected in the MUSE spectrum, at any wavelength corresponding to the redshift window where the AstroSat FUV detection would correspond to a foreground Lyman-alpha emitter. The only emission lines detected in the spectrum are Mg~{\sc ii} and [O~{\sc ii}] from AUDFs01, at $z\sim 1.42$.  
The flux density measured by AstroSat in FUV is $FUV =2.12\times10^{-18}$ erg.s$^{-1}$.cm$^{-2}$.A$^{-1}$. Assuming a Lyman-$\alpha$ equivalent width as weak as 2\AA, this translates into a predicted H$\alpha$ flux F(H$\alpha$) $> 2 \times FUV /8.7 \sim  5 \times10^{-19}$ erg.s$^{-1}$.cm$^{-2}$, this lower limit is calculated assuming case B, and an escape fraction of Lyman-alpha emission of 100\%. All escape fractions from LAEs reported in the literature are (much) lower, so the expected line fluxes in the optical are (much) stronger. 

Assuming that the detection in AstroSat is only due to the UV continuum of a foreground galaxy, the detected FUV magnitude, 25.84, would correspond to an SFR $\sim 3$~M$_{\odot}$~yr$^{-1}$, translating to an H$\alpha$ line flux of $\sim 3 \times 10^{-17}$ erg.s$^{-1}$.cm$^{-2}$. Given that the $3 \sigma$ detection threshold for an emission line in the MUSE deep data is $3.\times 10^{-19}$ erg.s$^{-1}$.cm$^{-2}$, we would be able to detect the optical nebular lines associated with an intervening galaxy.

 Another line of argument is to estimate the probability of detecting a LAE along the sightline. A cylinder of the Universe with a section of $1.6"$ and a length corresponding to $0.069 < z < 0.439$ would subtend a volume $\sim 0.6$~Mpc${^3}$. From previous GALEX studies (Fig 13 in Wold et al 2017), the number of galaxies with Lyman-$\alpha$ luminosities around $10^{41}$ erg.s$^{-1}$ is as low as $10^{-3}$ objects per Mpc$^3$ (brighter galaxies are even less numerous). So we would need a volume of 1000~Mpc$^3$ to have one chance alignment.

\subsection{Comparing with HST/F150LP detection}
It is important to compare our results with a previous search \cite{Sianaetal2007} for LyC leakers in the redshift range, $1.1 <z < 1.5$ with deep FUV imaging of the HST/XDF field in the ACS/SBC F150LP band. When we examined the F150LP image, we were unable to identify any source at the location of AUDFs01 confirming the previous conclusion\cite{Sianaetal2007}. 
The non-detection of AUDFs01 in the HST data can be understood from the sensitivity differences of the two images at the location of the source.

Although the read-noise is 0 e- in both SBC/MAMA and UVIT/CMOS, the dark current is higher in SBC/MAMA \cite{Teplitzetal2006,Timothy2016}; it is $\sim 1.2\times10^{-5}$~e$^{-}$~s$^{-1}$~pix$^{-1}$ (see ACS quick reference guide, cycle 15, 2005). In comparison, we have a dark current of $7.8\times10^{-7}$~e$^{-}$~s$^{-1}$~pix$^{-1}$. 
Accounting for the proper pixel scales and 
using the HST/F150LP weight image, we estimate the total background noise at the location of AUDFs01 and compare that with UVIT/F154W image. Using the total background noise, we compute the $3\sigma$ upper limit as 28.2 (within 0.5" radius), 27.4 (1.") and 27.2 (1.2") AB mag. The same for the F154W image are 28.7 (0.5"), 27.9 (1.") and 27.7 (1.2") AB mag. This shows that the background noise is higher in HST/F150LP image compared to F154W image. One can also think that UVIT is more sensitive to low surface brightness objects because it has a lower spatial resolution and thereby sampling more angular area per unit detector element.

Then we estimate the magnitudes of a few brighter sources common between the two images. For example, a source at 53.14472, -27.7854 (in the vicinity of AUDFs01) has 23.7 AB mag (aperture corrected) in F150LP image while it is 23.77 AB mag (aperture corrected) in F154W. Within errors, these are nearly the same. Using this comparison, we compute the expected flux within 1" radius for AUDFs01 in HST/F150LP band. We chose 1" as an intermediate aperture because 0.5" is too small for UVIT while 1.6" is large for HST. 
\noindent In F154W image, the background-subtracted flux for AUDFs01 is $1.16\times 10^{-4}$~ct~s$^{-1}$ within 1" (2.4 pix) radius. This would produce $0.008$~ct/s within 1." in F150LP image. The corresponding magnitude is 27.7 AB mag while the detection limit within 1" is 27.4 AB mag. This is, of course, a simple calculation (e.g., assumed wavelength coverage for two filters same); there could be a number of other reasons that we are not aware of. 

There are other differences between the two observations as well: the median exposure time at the location of AUDFs01 is $5262$~sec in F150LP while in F154W image, it is $63938$~sec. The blueward cut-off for F154W filter is 135 nm while that for F150LP is 145 nm. We have checked that between 135 - 145 nm, the F154W filter would contain about 12\% of the total flux for the best-fit CIGALE SED model of AUDFs01.

\begin{figure*}[h]
\rotatebox{0}{\includegraphics[width=\textwidth]{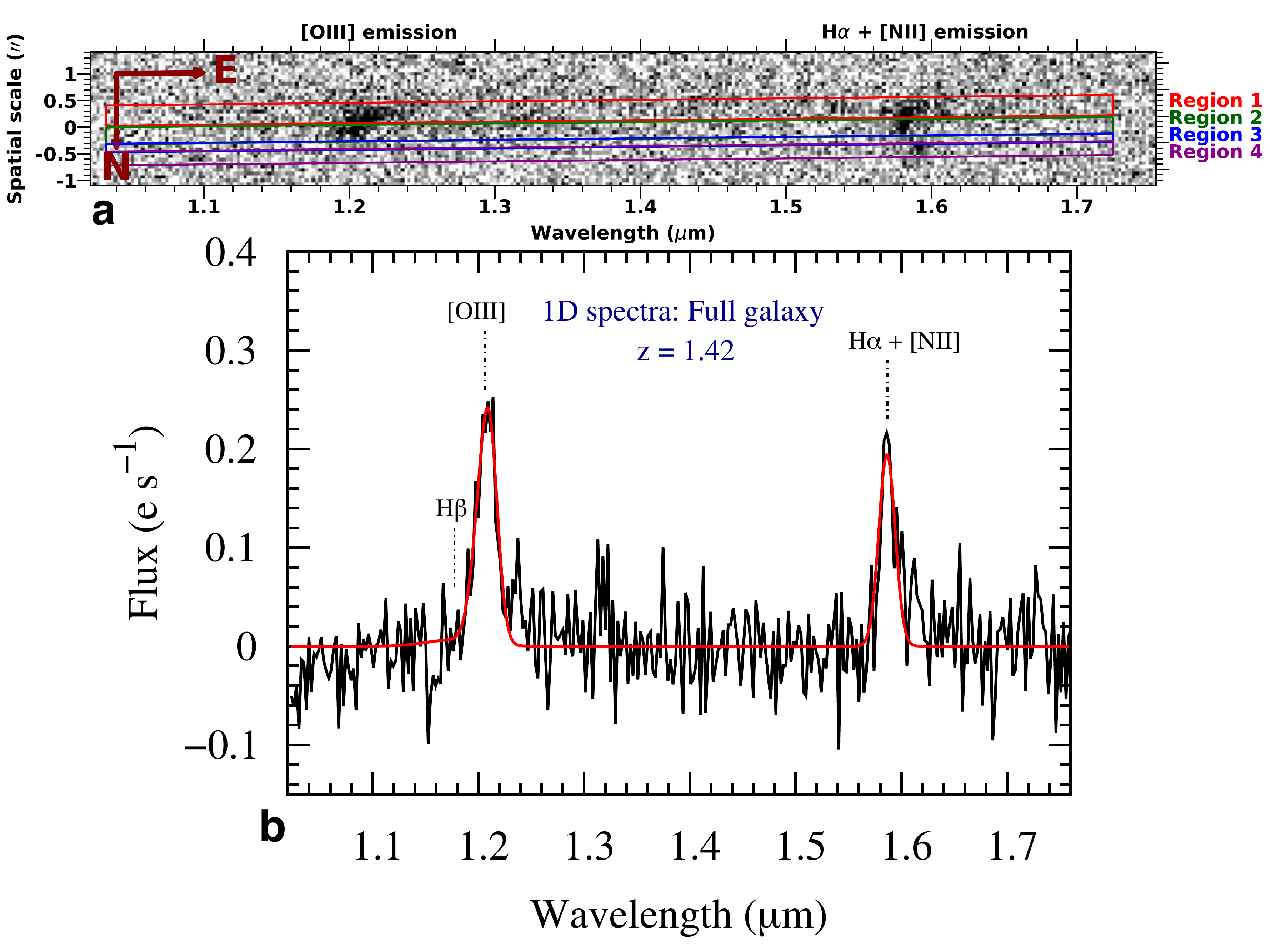}}
\caption{{\bf HST grism (G141) image} \cite{Momchevaetal2016}. {\textbf{a}}: Coloured rectangular regions (1,2,3,4) marked on the grism image are used to extract spectra for the clumps. North-East directions are marked on the grism image. {\textbf{b}}: 1D spectrum for the full galaxy. Red solid line represents the fitting of the spectrum. Redshift measurement is based on the fitting of H$\alpha$+[N~II] line alone.} 
\label{fig:EDFig1}
\end{figure*}

\subsection{Spectral analysis: Emission line mapping}
\label{spectraldata}
All four clumps (C1, C2, C3 and C4) in AUDFs01 can be identified in the HST optical (F435W, F606W, F775W, F814W, F850Lp), IR (F105W, F125W, F140W and F160W) and in the UV (F275W, F336W) bands. By analyzing the WFC3/F160W imaging data, we find that the centroids of the clumps C1 and C2 are aligned within a pixel while C3 and C4 are well separated. We use the clump centroids and the position angle (PA=130 deg) information to perform astrometric solution to the 2D direct image (3DHST id: 25778) in F160W. The F160W direct image is utilized to create an emission line mapping of the grism G141 image. We then extract the continuum plus the contamination subtracted grism spectra for each of four regions marked in Fig.~\ref{fig:EDFig1}. Both region~1 and region~2 contain two prominent broad emission lines, [O~{\sc iii}] and H$\alpha$. We choose the H$\alpha$ line to determine the spectroscopic redshift for the region~1 and region~2, which turn out to be at $z=1.420$ and $1.423$ for region~1 and region~2 respectively. There is no prominent ($5\sigma$ or above) emission in the grism data from the clump C4. But after masking region~1 (C1 and C2) and region~2 (C3), we found clear H$\alpha$ emission from C4. The H$\alpha$ emission from C4 is modeled with a Gaussian; the line flux is obtained to be $4.64 \times 10^{-17}$~erg~s$^{-1}$~cm$^{-2}$ and the redshift of C4 is found to be $z=1.415$. 
Such masking became possible only after creating an emission-line mapping which allowed us to relate pixel-to-pixel mapping between the grism and F160W direct image provided by the 3DHST observation. 
By fitting the contamination and continuum subtracted HST-Grism G141 spectra with Gaussian function, we have extracted the emission line properties of the full galaxy - in particular, the colour excess E(B-V) and Oxygen abundance. 

\subsection{Dust attenuation}
\label{sec:dust}

\subsubsection{UV slope method:}
The UV spectral slope $\beta$ (where $f_{\lambda} \sim \lambda^{\beta}$) is a sensitive indicator of the dust attenuation and hence can be used to estimate the continuum color excess E(B-V). The parameter $\beta$ can be formally defined\cite{Nordonetal2013} as: 

\begin{equation}
    \beta = -\frac{m(\lambda_{1})-m(\lambda_{2})}{2.5 \log({\lambda_{1}/\lambda_{2}})} -2
\end{equation}

\noindent where $m(\lambda_{1})$ and $m(\lambda_{2})$ are the AB magnitudes of the galaxy at rest-wavelength $\lambda_{1}$ and $\lambda_{2}$ respectively, where according to Calzetti et al. (1994), the rest-wavelength $\lambda \in [{1268 - 2580}]$\AA. Using the above relation, we find $\beta = -2.0\pm 0.26$ for AUDFs01. Then following the E(B-V) and $\beta$ relationship \cite{Reddyetal2018} appropriate for high-z galaxies, the continuum color excess is given by

\begin{equation}
    E(B-V)=\frac{1}{4.684}[\beta + 2.616] = 0.13 \pm 0.05
\end{equation}

It has been shown that star-forming galaxies on the main sequence (MS) relation in the redshift range $1 < z < 2.5$, follow a flatter relation between attenuation and UV slope than Meurer relation\cite{Meureretal1999}. Although AUDFs01 lies above the MS relation, we estimate the color excess assuming Calzetti reddening curve and the following relation

\begin{equation}
    E(B-V)=\frac{1}{k(1600\AA)}[1.26 \beta + 3.90] \simeq 0.14 \pm 0.03
\end{equation}

Our best-fit CIGALE SED model (discussed in Sec.~\ref{sec:cigale-Bpass}) yields E(B-V)=0.15 in close agreement with that derived from the UV spectral slope $\beta$.

\subsubsection{Balmer decrement method:}
\label{sec:Balmer}
The grism G141 is ideal to capture two of the prominent emission lines such as [O~{\sc iii}]~5007 and H$\alpha$ for star-forming galaxies in the redshift range $1.1 < z < 1.5$. In this work, we utilize the emission line data from grism G141 observed under the 3DHST program \cite{Momchevaetal2016}. 
The nebular colour excess can be estimated using the method of Balmer decrement\cite{OsterbrockFerland2006} assuming a case B recombination, temperature $T=10^4$~K and electron density $n_e = 100 cm^{-3}$ as: 

\begin{equation} 
E(B-V) = 1.97 \log\big{[}\frac{(H{\alpha}/H{\beta})_{obs}}{2.86}\big{]},
\label{eq:balmer}
\end{equation}

\noindent where H$\alpha$ and H$\beta$ are the observed line fluxes. In the grism G141, the H$\alpha$ and [N~{\sc ii}] lines are blended. We use a prescription based on equivalent width\cite{Sobraletal2012} to extract the [N~{\sc ii}]~6583~$\AA$ line intensity from the blended H$\alpha$ line. Further, we have extracted the NII~6548~$\AA$ line flux by applying the atomic transition probabilities\cite{Osterbrock1989} such that the ratio of [N~{\sc ii}]6583/[N~{\sc ii}]6548 = 2.93. Using the measured fluxes in Eq.~\ref{eq:balmer}, we obtain $E_{nebular}(B-V)=0.48 \pm 0.40$. Note the large uncertainty in the determination of nebular color excess comes from the poor $S/N$ of the $H{\beta}$ measurement in the grism G141. In fact, this has been one of the major uncertainties in estimating the dust extinction using Balmer decrement at this redshift range. Moving ahead, we use the Calzetti relation to estimate the continuum color excess $E_{star}(B-V)=0.44 E_{nebular}(B-V) = 0.21$. Since continuum color excess estimated using the UV $\beta$~slope method and PCIGALE SED fitting are in close match with each other, we prefer to use the value from UV $\beta$ slope i.e., $E(B-V)=0.13$ throughout this paper unless mentioned otherwise.

The [O~{\sc ii}] line flux is extracted from the MUSE archival data \cite{Inamietal2017}. It is extinction corrected and slit loss is null in the image slicing mechanism used in the MUSE IFU \cite{Baconetal2017}.
All the emission line fluxes from the grism and MUSE are corrected for the foreground Galactic extinction \cite{Schlegeletal1998} of $A_{V}=0.019$~mag and further corrected for the internal extinction with a continuum color excess $E(B-V)=0.13$ derived from the UV $\beta$ slope. Finally, the line luminosities are derived using a luminosity distance of $D_{L}= 10194.6$~Mpc and these are presented in Table~\ref{tab:tab1}.

\subsection{Oxygen abundance}
\label{sec:Oxygen}
The Oxygen abundance in the galaxy is estimated following the N2 method\cite{PettiniPagel2004}: 
\begin{equation}
    12 + \log[O/H] = 8.90 + 0.57 \times N2 =7.99,
\end{equation}

\noindent where $N2 = \log {[N~{\sc II}]~6583}/H{\alpha}$=-1.59. In terms of solar metallicity, [O/H] $\sim 1/5$ solar -- indicating a metal-poor galaxy with Z=0.0042. In this calculation, we used
[N~{\sc II}]~6583 and H${\alpha}$ fluxes derived following the method of equivalent width \cite{Sobraletal2012}. 

We have also estimated Oxygen abundance using line fluxes from 3DHST catalogue fluxes. In all cases, the N2 method gives $Z \sim 0.004$. 

In addition, we have also used the O3N2 method\cite{PettiniPagel2004} following the relation 
$$12 + \log[O/H] = 8.73 - 0.32 \times O3N2 = 7.92,$$

\noindent where $O3N2 = \log{ \frac {[O~{\sc III}]/H{\beta}} {[N~{\sc II}]/H{\alpha}}  } =2.52.$ In terms of solar metallicity, [O/H] $\sim 1/6$ solar -- indicating again a metal-poor galaxy with Z=0.0034. Since O3N2 relation is valid only for $O3N2 \le 1.9$, we restrict to N2 method in this work.

\begin{figure*}
\begin{center}
\rotatebox{0}{\includegraphics[width=0.9\textwidth]{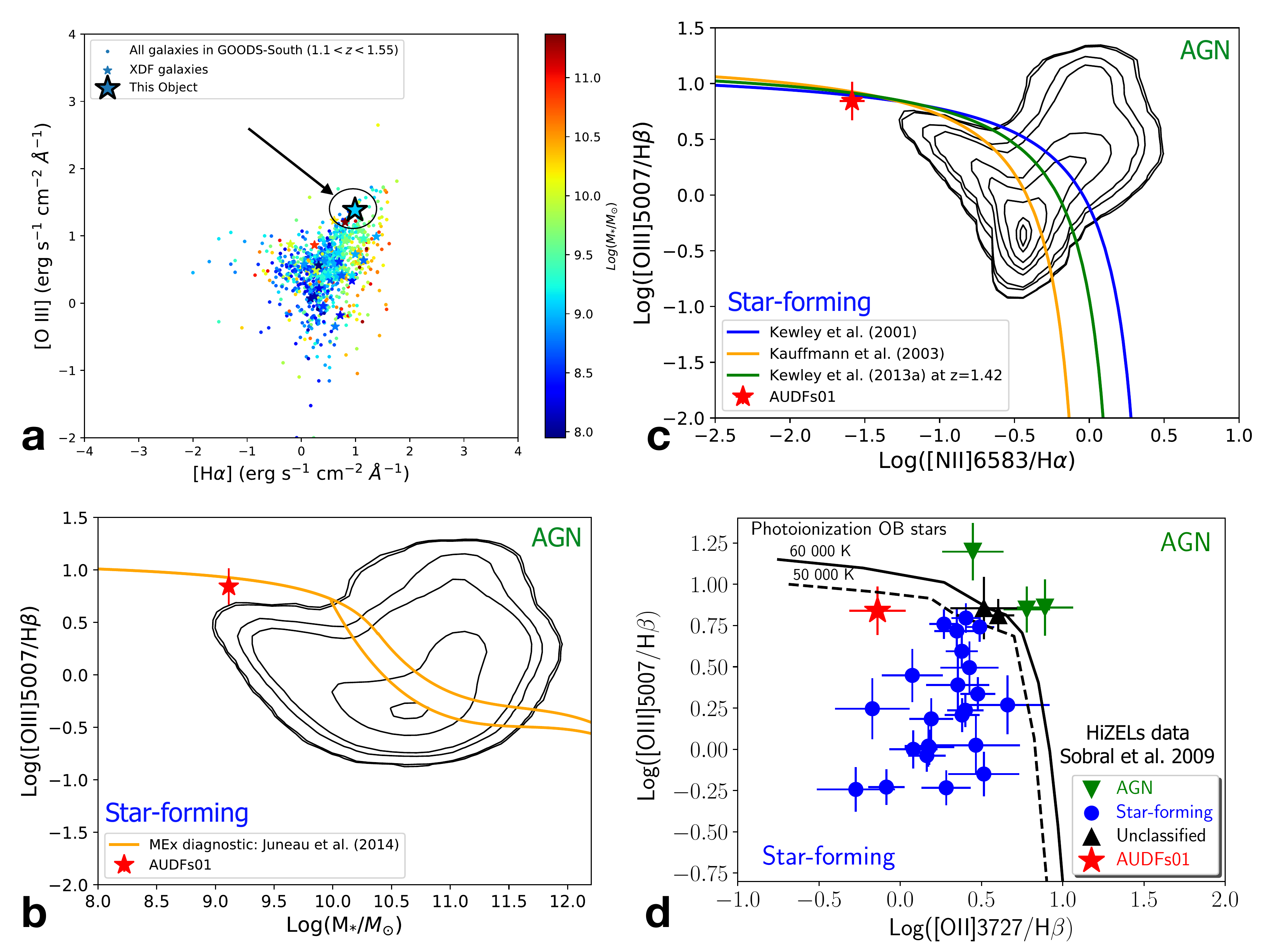}}
\caption{{\bf SF-AGN diagnostic diagram:} {\bf a}: location of the clumpy galaxy AUDFs01 on the H${\alpha}$ - O~{\sc iii} plane. The line fluxes are measured from HST grism G141 data\cite{Momchevaetal2016}. AUDFs01 being the only galaxy having highest O~{\sc iii} flux in the XDF region; the color bar indicates the stellar masses of the galaxies. 
{\bf b, c}: Mass Excitation and BPT diagram using the SDSS galaxies. {\bf d}: location of AUDFs01 on the Sobral et al. 2009 plot. The line ratios for all galaxies except AUDFs01 are taken from z-COSMOS survey\cite{Sobraletal2009} at $z\sim0.84$. The error bars represent $1\sigma$ uncertainties on the flux measurements. }
\label{fig:EDFig2}
\end{center}
\end{figure*}

\begin{figure*}
\rotatebox{0}{\includegraphics[width=1.0\textwidth]{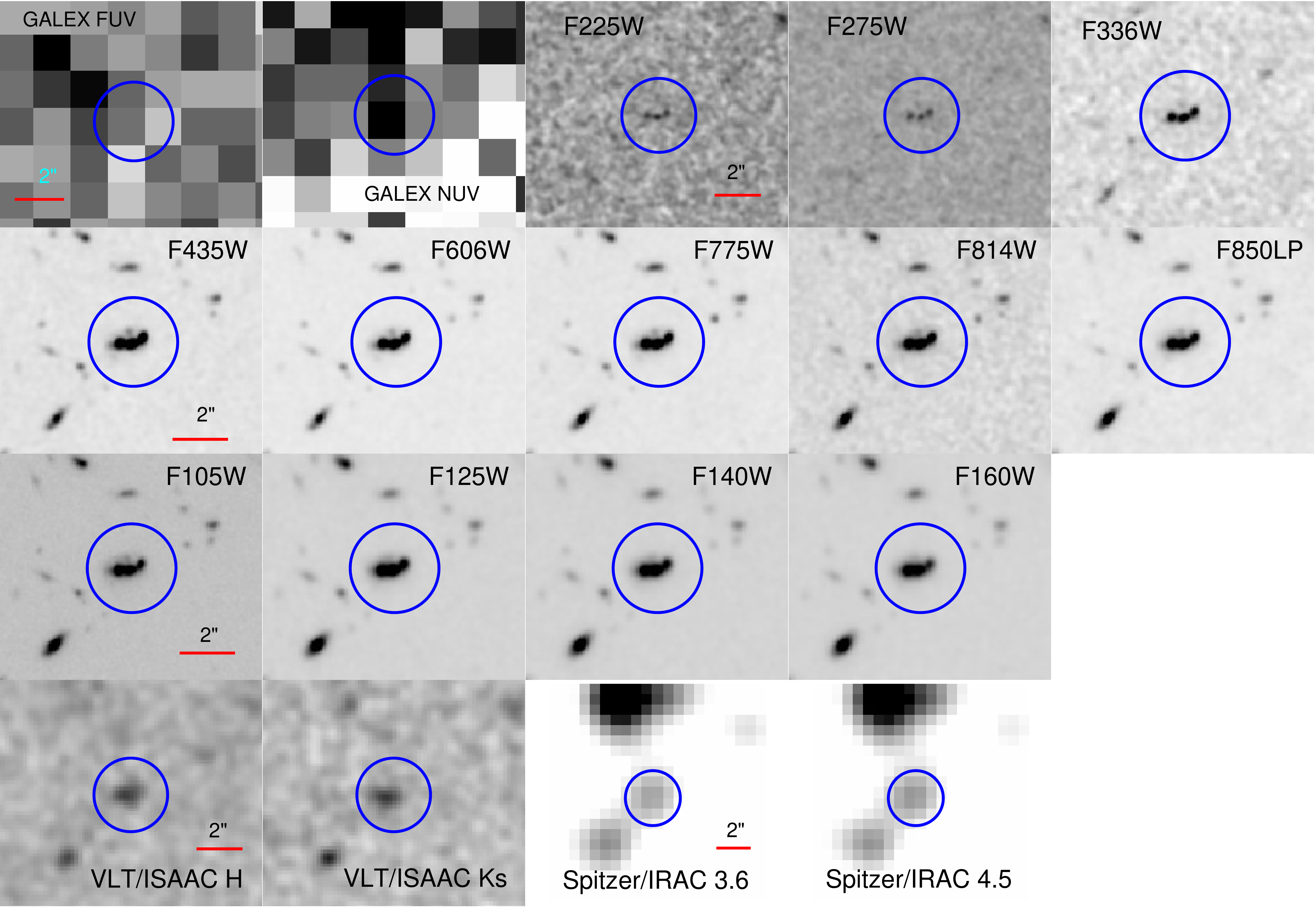}}
\caption{{\bf Postage stamp images:} GALEX (FUV, NUV), HST (UV, Optical, IR), VLT/ISAAC (H, Ks), Spitzer/IRAC (3.4, 4.5 micron)-bands. The radius of the blue circle in each panel is 1.6".}
\label{fig:EDFig3}
\end{figure*}

We have used the emission lines to construct the BPT\cite{BPT1981} and mass-Excitation diagram \cite{Jeneauetal2014} for the galaxy (Fig.~\ref{fig:EDFig2}). It is inferred that the galaxy does not host an obvious active galactic nucleus (AGN); it belongs to the star-forming galaxies (SFG) in the BPT. The galaxy is not detected in the 2Ms or the 7Ms Chandra source catalog \cite{Luoetal2017} - confirming the non-AGN nature. The same has been confirmed when we compare AUDFs01 with the z-COSMOS SFGs at z=0.84 with [O~{\sc iii}]/H$\beta$ and [O~{\sc ii}]/H$\beta$ ratios \cite{Sobraletal2009}.
When searched in the ALMA deep fields, within an RMS of 35 microJy, there is no detection in the 1.3mm continuum \cite{Dunlopetal2017, Francoetal2018} - indicating less of a molecular gas content in parts of the galaxy. From this, we obtain an upper limit of $M_{gas} < 10^{9} M_{\odot}$ - in close agreement with our estimate ($M_{gas}=2.2\times 10^{8}$~M$_{\odot}$) from SED fitting.

\begin{table*}
\caption{{\bf Emission-line flux and luminosities.} Col2: line flux as measured in the HST grism G141; [O~{\sc{ii}}] is from MUSE catalogue. col3: line fluxes after foreground dust plus internal extinction correction using Balmer decrement. Col4: same as col3 but internal extinction (E(B-V)=0.13) due to UV beta slope; the value of H$\beta$ (marked bold-face in the bracket) is what would be expected as per the internal Balmer decrement given the measured H$\alpha$. Col5: line luminosity following UV beta slope. }

\begin{tabular}{cccccccc}  \hline\hline
Line & flux & flux1$_{corr}$ & flux2$_{corr}$ & L$_{\beta}$\\
         &  {EW method} & (Balmer decrement) & (UV $\beta$) & \\ 
       & ($\times 10^{-17}$ cgs unit) &($\times 10^{-17}$ cgs unit)& ($\times 10^{-17}$ cgs unit) &($\times 10^{41}$~erg~s$^{-1}$)  &\\
\hline
O~{\sc ii}~3726.0 &$0.48\pm0.009$ &$6.64\pm 0.12$ & $2.44\pm 0.045$ &$3.04 \pm 0.06$\\
O~{\sc ii}~3729.0 &$0.69\pm0.012$ &$9.52\pm 0.16$ & $3.50\pm 0.06$ &$4.36 \pm 0.07$\\
H${\beta}$~4861.0 & $1.86\pm 0.78$ & $14.63\pm 6.1$ & $6.7 ({\bf 8.2})\pm 2.7$ &$8.30\pm 3.4$\\
O~{\sc iii}~4959.0 &$5.53\pm  0.40$ & $41.7\pm 3.0$ &$19.3\pm1.4$ &$24.0\pm 1.7$ \\
O~{\sc iii}~5007.0 &$16.59\pm 1.20$ &$122.5\pm8.8$& $57.2\pm 4.1$ &$71.2\pm5.1$ \\

N~{\sc ii}~6548 & {$0.08\pm 0.007$} &$0.36\pm0.03$ &$0.20\pm0.02$ &$0.25\pm 0.02$\\

H${\alpha}$~6563.0 & {$9.33\pm 0.96$} & $41.3\pm 4.2$ & $23.46\pm2.4$ &$29.2\pm 3.0$\\

N~{\sc ii}~6583  &{$0.24\pm 0.022$} & $1.06\pm0.09$ & $0.60\pm0.05$ &$0.75 \pm 0.07$\\

\hline
\end{tabular}
\label{tab:tab1}
\end{table*}

\subsection{Multi-wavelength SED modelling} 
\label{sec:cigale-Bpass}
{\bf CIGALE modelling:}

We have constructed a multi-wavelength SED from FUV to IR ($1300 - 45000 \AA$) for the full galaxy (Fig.~\ref{fig:fig3}). For the F154W and N242W, we have used a fixed circular aperture of radius 0.9" and applied the method of growth curve correction to estimate the total flux of the galaxy. The same procedure was followed to estimate fluxes in VLT/ISAAC H, Ks and Spitzer/IRAC $3.6$ and $4.5$~{$\mu m$} bands; in that we have obtained growth curve correction factor in each of these bands. In case of HST also, we have used a fixed aperture of 0.9" to calculate the flux and applied the aperture correction - this has been followed throughout all bands of HST used in the final SED construction. In each case, we examine the region around the object carefully by eye to ensure that there is no contamination from other sources. The error on the measured fluxes are estimated using the available archival weightmaps. Our measurement of fluxes are given in Table.~\ref{tab:tab3} and these values are in good agreement with 3DHST measurements\cite{Momchevaetal2016}; for example, our F775W flux corresponds to $24.16\pm 0.05$ AB mag whereas 3DHST catalogue value is $24.19\pm0.03$ AB mag.

The physical properties of AUDFs01 have been derived by fitting the stellar population model with nebular lines using PCIGALE \cite{Boquienetal2019} as well as using binary stellar population BPASS \cite{Eldridgeetal2017}. In the PCIGALE modeling, we use BC03 stellar population library\cite{BC03} with exponentially declining star formation histories and late bursts. We employ a Salpeter Initial Mass Function (IMF)\cite{Salpeter1955} with lower and upper mass cutoffs at 0.1 and 100 M$_{\odot}$ respectively. The metallicity values used are $0.0004, 0.004,0.008,0.02$; note the metallicity from our emission line measurements is $Z\sim 0.004$. The colour excess was varied from E(B-V)=0.11 - 0.22. This range was chosen from our estimates of color excess derived based on UV spectral slope (0.13) and Balmer decrement (0.21). In the SED modeling, we followed the Calzetti relation $E_{star}(B-V)=0.44 E_{nebular}(B-V)$; although there is a considerable debate ongoing for high redshift galaxies\cite{Reddyetal2016}. We apply Calzetti extinction law \cite{Calzettietal2000} for dust modeling. The dust attenuation curve has a UV bump at 2175~\AA~ with amplitude $\sim 1/3$ of that of the MW bump and the overall power-law slope of the curve is fixed at $n=-\delta +0.75 = 1.25$, where $\delta$ is the slope deviation. This slope is similar to the slope of the SMC extinction curve, which may be more appropriate for high redshift SFGs\cite{Salimetal2018}. The slope deviation ($\delta$) closely matches with the relation \cite{Salimetal2018}: $\delta = -0.38+ 0.29\times(\log M_{*} - 10.)$. Various modules and their parameters used in the SED modeling are given in Table~\ref{tab:tab2} and their detailed description can be found here \cite{Boquienetal2019}.
For the CIGALE models that are shown in Fig.~\ref{fig:fig3}, we have fixed $Z=0.004$, IMF to Salpeter, Ionization parameter to -3.0; E(B-V) factor to 0.44 and some of the dust parameters as in Table.~\ref{tab:tab2} but we have allowed E(B-V) to vary from 0.11 to 0.22 in steps of 0.01; $f_{\rm esc}$ was varied from 0. - 0.8, in steps of 0.1. The best-fit PCIGALE model is chosen based on the minimum $\chi^2$ value as well as one that closely matches the ratio of EW of the observed lines (e.g., H$\alpha$, [O~{\sc iii}]) from HST grism G141. 
The best-fit PCIGALE model yields E(B-V)=0.15 and $f_{\rm esc} =0.5$. Since the best-fit continuum color excess E(B-V) is close to one from the UV $\beta$ slope method, we have also obtained a CIGALE model with E(B-V)=0.13 (UV slope) and for the sake of completeness with E(B-V)=0.2 (Balmer decrement), see Fig.~\ref{fig:fig3}. In both cases, the modelling was done exactly in the same way as for the best-fit but with E(B-V) fixed. For E(B-V)=0.13 case, the model yields $f_{\rm esc} =0.45$. The best-fit CIGALE model shows an emission line He~I 635~\AA~ falling in the observed FUV band. We have estimated the total flux by integrating the best-fit model over the F154W band with and without the 635~\AA~ line. We found the line flux contributes to 2\% of the total flux without the line and conclude that this alone could not have boosted the observed FUV flux.  

\medskip
{\bf BPASS modeling:}

As an independent method, we used simple stellar population models to estimate the age and metallicity range in which our observations could be reproduced. These assume that all stars are born at the same time and distributed according to their IMF. In particular, we used BPASS\cite{Eldridgeetal2017} models to create a grid (Table~\ref{tab:tab2}) that allowed us to explore several different spectral shapes. Additionally, we used these spectra to simulate a photoionized region using the software Cloudy\cite{Ferlandetal2013}. Most parameters were maintained fixed, varying only the stopping condition of the code, in our case determined by column density. To obtain results in which some ionizing radiation escapes it was necessary to probe low column density values ($10^{16}$, $10^{17}$, $10^{18}$ cm$^{-2}$).
Besides, we explored all of the models available in Starburst99 \cite{Leithereretal1999}. The results were consistent with the ones found using BPASS spectra, indicating a low metallicity ($Z/Z_{\odot}$ $\sim$ 0.004) and young stellar age ($<$ $5\times10^6$ years). As a test, we computed the ratios $f_{\lambda, 600}/f_{\lambda, 1500}$ and $f_{\lambda, 900}/f_{\lambda, 1500}$ for all the analyzed models and confirmed the possibility of obtaining the observed ratios using any of the models in the mentioned metallicity and age ranges. It is important to note that in this analysis, we used the intrinsic stellar spectra provided by the models to avoid making as many assumptions as possible.

In Fig.~\ref{fig:fig3}, we show the chosen BPASS spectra after being run through the photoionization code Cloudy \citep{Ferlandetal2013}. The chosen parameters are $Z = 0.004$, age = $5\times10^6$~yr, ionization parameter $U=-1.5$, and $\log(N_{H}=17$ cm$^{-2}$. This choice was made based on the closeness to the Astrosat observations, which happen to be the same main parameters as those found by a Cigale fitting. To obtain a more realistic result, we applied some dust extinction and IGM attenuation. For the first one, we used the extinction law by Reddy+2016. This law is only valid down to 950\AA, however, since there is no reliable extinction law below this limit, we decided to extrapolate it to lower wavelengths as an approximation. For the second one, we used Monte-Carlo simulation \cite{InoueIwata2008,Inoueetal2014} to find the appropriate range of IGM attenuation. IGM transmission, being extremely stochastic, depends heavily on the line of sight observed. Our analysis of the transmission values at 600~\AA ~and intermediate redshift reveals a bimodal behaviour (See Fig.~\ref{fig:fig4}, right panel). The transmission distribution allows only values either close to $0$ or close to $1$ at lower wavelengths, high or low, with 50\% of the lines of sight having a transmission above 80\%. Since ionizing photons were detected, it is reasonable to assume that the transmission is high along our particular line of sight.

\begin{table*}
\small\addtolength{\tabcolsep}{-3pt}

\begin{flushleft}
\begin{tabular}{ccccccc}  \hline\hline
Module     & physical parameters & parameter range (best fit - bold face)\\
       & &  &\\

\hline
Sfh2exp    & $\tau_{main}$ (Myr) & {{\bf 50}, 100, 200, 400, 700, 1000, 2000, 4000}    \\

                 &$\tau_{burst}$ (Myr) & 50, {\bf 100.},150.\\
                 &$f_{burst}$ & {0.01,0.02,0.04,{\bf 0.06},0.08,1.0}\\
                 &$Age_{main}$ (Myr) & {20.,50.,{\bf 100},200,300,500,600,700, 1000,2000, 4000}\\
                 &$Age_{burst}$ (Myr) & {\bf 2.0},3.0,4.0,5.0,10.0\\
 \hline                
BC03  & IMF & 0 ({\bf Salpeter})\\
           & Metallicity & 0.0004, {\bf 0.004},0.008, 0.02\\
 \hline          
Nebular & Ionization parmeter (logU) & -3.0\\
             &fraction of LyC photons escape $f_{esc}$  & 0.0,0.1,0.2,0.3,0.4,{\bf 0.5},0.6,0.65,0.7,0.75,0.8\\
             &fraction of LyC photons absorbed $f_{dust}$ & 0.0,0.05,0.1,{\bf 0.15},0.2\\
             &Line-width (km/s) & 300.0\\
 \hline            
Dustatt-Calzetti & E(B-V)$_{young}$ (mag) & 0.11,0.12,0.13,{\bf 0.15},0.17,0.19,0.2,0.22 \\
                           &E(B-V)-{factor} & {\bf 0.44},0.6,0.9,1.0\\
                           &uv-bump-wavelength (nm) & 217.5 \\
                           &uv-bump-width (nm) & 35.0\\
                           &uv-bump-amplitude & 1.0 \\
                           & ($\delta$) modifying the attenuation curve & -0.5\\
                           
dl2014   & $\alpha$ (as in $dU/dM \propto U^{\alpha}$) & 2.0\\
\hline
restframe-parameters & beta-calz94 & True\\
                                    &Dn400 & True\\
                                    &IRX & True\\
                                    &EW$_{lines}$ (nm) & 372.7/1.0, 486.1/1., 500.7/1. 656.3/1., 658.3/1.\\ 
                                    &Luminosity$_{filters}$ & F154W, N242W, V$_{B90}$\\
                                    &Color$_{filters}$ & FUV-NUV, NUV-r$^{\prime}$\\

\hline
\hline
\end{tabular}
\end{flushleft}
\begin{center}
    \title{BPASS parameter space:}
    \begin{tabular}{l|l} 
      \textbf{Parameter} & \textbf{Allowed values}\\
      \hline
      Age [yrs] & $5\times10^6$, $10^7$, $5\times10^7$, $10^8$, $5\times10^8$\\
      Metallicity [$Z$] & $10^{-5}$, $10^{-4}$, 0.002, 0.004, 0.008, 0.020, 0.040\\
      Column density [$N_{H}$ cm$^{-2}$] & 10$^{16}$, 10$^{17}$, 10$^{18}$\\
      Model & single, binary
    \end{tabular}
  \end{center}

\caption{{\bf SED fitting parameters for CIGALE and BPASS}. The best-fit parameters for cigale modelling are indicated by the bold-face letters. Dn4000 represents the ratio of the average flux density in two two narrow bands, $3850 - 3950$~\AA and $4000 - 4100$~\AA. IRX refers to the infrared excess}
\label{tab:tab2}
\end{table*}

\par
{\bf Clump SED modeling:}

We have used 11 pass-bands of HST UV/optical/IR to construct SED for each clump, with wavelength ranging from $2750$ \AA ~to $16000$ \AA, containing the $4000$ \AA~ break. For this purpose, we have used all HST images with 0.06" resolution. The flux of each clump has been extracted by placing a square box with a fixed size of 0.3" around each centroid of C1, C2, C3 and C4 in F606W image. Then we have made measurements away from the clumps, by placing the box at 7 different locations including intra-clump as well as the outskirts of the galaxy. We take the mean of these measurements and subtract it from C1, C2, and C3 to determine their fluxes. For C4, we made 3 such measurements - left, right and top of the C4 centroid. The mean of these three measurements is subtracted from C4, to determine the C4 flux. This process is repeated identically for all 11 pass-bands. These fluxes are corrected for the Galactic extinction \cite{Schlegeletal1998} and the magnitudes of the clumps are presented in Table~\ref{tab:tab3} along with the full galaxy. We then derive the photometric redshift of each clump by modeling their SED using  EAZY\cite{Brammeretal2008}. In this fitting, we take into account of the IGM absorption \cite{Madau1995} and let the redshift vary from z=0.1 to 6.0. The photometric redshifts obtained are: C1 at $z=1.39$; C2 at $z=1.7$;  C3 at $z=1.42$, and C4, $z=1.40$. Our photometric redshift estimates are in close agreement with those derived from the grism G141 spectrum extracted for each clump. We then perform CIGALE modeling to derive physical parameters of the clumps using the best-fit SED parameters of the full galaxy e.g., we use E(B-V)=0.15, Z=0.004. We found the masses of the clumps C1, C2 and C3 to be similar to each other while C4 is $\sim 100$ times lower than the rest.

\begin{table*}[ht]
\centering
\caption{{\bf Magnitudes of the galaxy AUDFs01 and its clumps} at different passband. All magnitudes are aperture and foreground dust corrected.}
\begin{tabular}{cccccccc}  \hline\hline
Filters & Full galaxy & C1 & C2 & C3 & C4\\
         & AB mag & AB mag & AB mag & AB mag & AB mag &\\ 
\hline

F154W & $25.84\pm0.34$ &  & & &\\

N242W & $25.60\pm0.10$ & & & &\\

F225W & $25.45\pm 0.13$ & & &\\

F275W & $24.73\pm0.12$ & $26.74\pm 0.04$ &$26.76\pm 0.049$ &$26.88\pm 0.05$ & $28.94\pm 0.23$\\

F336W &$24.19\pm0.05$ & $26.25\pm 0.042$ & $26.16\pm 0.049$ &$26.47 \pm 0.05$ & $28.75 \pm 0.23$\\

F435W &$24.21\pm0.07$ &$26.18\pm 0.032$ & $25.96\pm 0.032$ &$26.41 \pm 0.023$ &$28.85 \pm 0.09$ \\

F606W & $24.20\pm 0.10$ & $26.12\pm 0.038$ & $25.76\pm 0.039$ &$26.35\pm 0.032$ & $28.94\pm 0.087$\\

F775W &$24.16\pm  0.05$ & $26.09\pm  0.045$ & $25.72\pm0.042$ & $26.38\pm 0.023$ &$29.04\pm0.11$ \\

F814W &$24.14\pm 0.20$ &$26.04\pm 0.055$ & $25.68\pm0.049$ & $26.33\pm0.044$ &$29.00\pm 0.26$ \\

F850LP & $23.77\pm 0.04$ &$25.86\pm 0.05$ &$25.59\pm 0.047$ & $26.22\pm 0.038$ & $28.88\pm0.16$ \\

F105W & $23.61\pm 0.21$ & $25.49\pm0.29$ &$25.40\pm 0.25$ & $25.93\pm 0.25$ &$28.56\pm 0.51$\\

F125W  &$23.40\pm 0.30$ &$25.46\pm 0.36$ & $25.34\pm 0.22$ & $25.91\pm 0.21$ &$28.56\pm0.70$\\

F140W  &$23.41\pm 0.36$ &$25.46\pm 0.36$ & $25.28\pm 0.23$ & $25.89 \pm 0.22$ &$28.58\pm0.74$\\

F160W  &$23.42\pm 0.20$ &$25.47\pm0.36$ & $25.24\pm 0.22$ & $25.88\pm 0.22$ &$28.61\pm0.74$\\

VLT/ISAAC H  &$23.74\pm 0.37$ &         &  & &\\

VLT/ISAAC Ks  &$23.95\pm 0.26$ &         &  & &\\

Spitzer $3.5 \mu m$  &$23.53\pm 0.13$ &         & & &\\

Spitzer~$4.5 \mu m$  &$23.29\pm 0.14$ &          & & &\\

\hline
\end{tabular}

\label{tab:tab3}
\end{table*}

\subsection{Ionizing photon production and escape fraction} 
The amount of leaking ionizing radiation is estimated from the observed number of LyC photons ($N_{LyC}^{obs}$) and that are intrinsically produced in the system ($N_{LyC}^{int}$). Assuming case~B  recombination, temperature $T=10^4$~K and electron density $n_e = 100$~cm$^{-3}$, we estimate the number of LyC photons that are capable of ionizing hydrogen atoms using the extinction corrected H$\alpha$ luminosity \cite{Osterbrock1989,Bergvilletal2006}:

\begin{equation}
N_{Lyc}^{non-esc} =  \frac{\alpha_{B}(H^{0},T_e)}{\alpha_{H_{\beta}}^{eff} (H^{0},T_e)} \frac{\lambda_{{H\beta}}/\lambda_{{H\alpha}}}{j_{H{\alpha}}/j_{H{\beta}}} \frac{L_{H{\alpha}}}{h \nu_{H{\alpha}}} = 7.28 \times 10^{11}L_{H{\alpha}}
\label{eq:non-esc}
\end{equation}

\noindent where $L_{H{\alpha}}$ is the luminosity of the H$\alpha$ emission line, $j_{H{\alpha}}/j_{H{\beta}}$ denotes the intrinsic ratio of H$\alpha$ and H$\beta$ line intensity, $\alpha_B$ and $\alpha_{H\beta}^{eff}$ are the Case B recombination rate and H$\beta$ emissivity \cite{Osterbrock1989} respectively. Other symbols have the usual meaning. The extinction-corrected H${\alpha}$ luminosity is estimated to be $(29.2 \pm 3.0) \times 10^{41}$~erg~s$^{-1}$ corresponding to $N_{LyC}^{non-esc} = (2.12 \pm 0.22) \times 10^{54}$~s$^{-1}$. A similar relation could be obtained for the $H\beta$ luminosity.

We estimate the rate of LyC photons that escape the galaxy directly from the FUV flux measured at F154W filter by AstroSat which probes $537 -723$ \AA ~rest-frame photons for this galaxy. The rest-frame luminosity of the galaxy in F154W is $1.68 \pm 0.45 \times 10^{43}$~erg~s$^{-1}$ corresponding to about $N_{LyC}^{emit}=0.54\pm0.14 \times 10^{54}$ LyC photons/sec ({\it{a lower limit, since we exclude the number of ionizing photons that are in the bluer part of N242W filter}}). Considering a transparent IGM ($\tau_{IGM}=0$) and the median value of $\tau_{IGM}=0.29$ along our line of sight, at this redshift, the escape fraction would be given by

\begin{align}
f_{\rm esc} = 
\frac{N_{LyC}^{emit} e^{\tau_{IGM}}}{N_{LyC}^{emit} e^{\tau_{IGM}} + N_{LyC}^{non-esc} }  = { 0.20 \pm 0.027} \label{eq:fesc1}\\
  ={ 0.25 \pm 0.029}
  \label{eq:fesc2}
\end{align}

\begin{figure*}[!h]
\begin{center}
\rotatebox{0}{\includegraphics[width=1.0\textwidth]{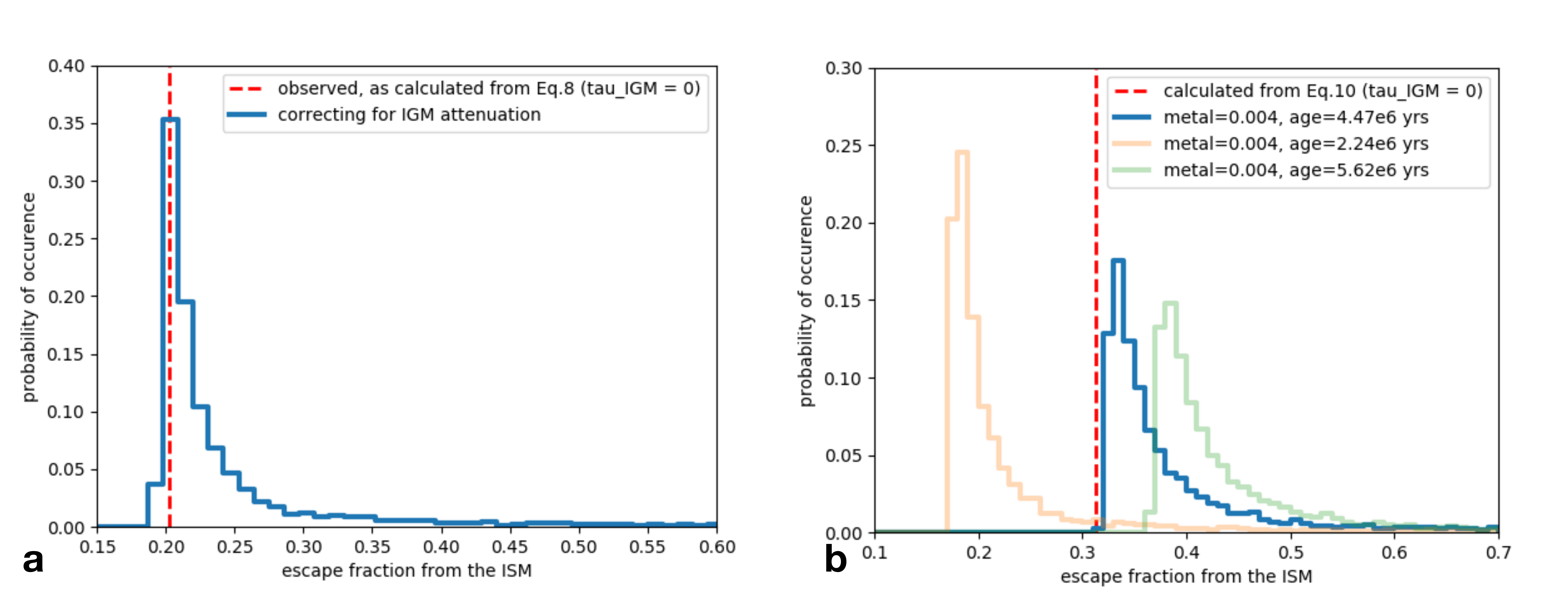}}
\vspace{-0.5cm}
\caption{{\bf Distributions of LyC escape fractions.} {\bf a:} For the first method, calculating the LyC escape fraction from the H$\alpha$ luminosity following Eq.~\ref{eq:fesc1}. 
{\bf b:} Following Eq.~\ref{eq:fesc600}, for the best fit BPASS model with metallicity $Z=0.004$, and an age of the stellar burst of $\sim 4.5\times10^6$ years. Other BPASS models, varying ages, are shown with faded lines for comparison.
On both panels, the vertical dashed line shows the value of the escape fraction assuming a transparent IGM (following Eq.~\ref{eq:fesc1} and Eq.~\ref{eq:fesc600}).}
\label{fig:EDFig4}
\end{center}
\end{figure*}

In order to correct for IGM attenuation, we divide $f_{\rm esc}$ from Eq.~\ref{eq:fesc1} by the IGM transmission along the line of sight. But this peculiar attenuation is not known, only the statistics of the IGM attenuation is known (See Fig.~\ref{fig:fig4}). The shape of the distribution of escape fractions corresponding to the distribution of non-zero IGM transmission is shown in Fig.~\ref{fig:EDFig4}(a).
Since we detect AUDFs01 in the FUV filter, the IGM transmission along our line of sight is non-zero, and we keep only the non-zero transmission of the distribution. The probability of occurrence of each escape fraction bin is normalised to the total number of non-zero transmissions. 
{\it The distribution of escape fraction peaks around 0.2 and most of the sightlines (90\%) have $0.20 < f_{\rm esc} < 0.31$} (Fig.~\ref{fig:EDFig4}). 

In the PCIGALE modeling \cite{Boquienetal2019}, nebular emission is modeled using nebular templates \cite{Inoue2011} that are generated using CLOUDY\cite{Ferlandetal2013} given a range of metallicity and ionization parameters. In that, the line luminosities are rescaled by the ionizing photon luminosity. The two primary factors that affect the ionization rate of the surrounding gas are $f_{\rm esc}$ -- a fraction of the LyC photons that simply escape and $f_{\rm dust}$  -- a fraction of LyC photons that are being absorbed or scattered by the dust. In the PCIGALE modeling, both are free parameters and $f_{\rm esc}$ was varied from 0.0 to 0.8 (Table.~\ref{tab:tab2}). While doing the SED modelling, we have always aimed at reducing the number of parameters to be constrained. The relevant parameter range that was varied to obtain models presented in Fig.~\ref{fig:fig3} is mentioned in sec.~\ref{sec:cigale-Bpass}. {\it The best-fit pcigale model yields a value of E(B-V)=0.15, $f_{\rm dust} =0.15$ and $f_{\rm esc}=0.5$.}

In a third method, the LyC escape fraction is derived by comparing the observed LyC flux to the intrinsic one expected from stellar population models using the following relation \cite{Leitetetal2013,Izotovetal2016a}:

\begin{equation}
f_{\rm esc}= \frac{(f_{600}/f_{1500})^{obs}}{(f_{600}/f_{1500})^{int}} e^{\tau_{IGM}} 10^{-0.4 A_{\lambda}(1500) }
\label{eq:fesc600}
\end{equation}

\noindent where $A_{\lambda} = k_{\lambda} E(B-V)$; $f_{1500}$ and $f_{600}$ are the flux densities at rest-frame wavelength. In writing the above equation, it has been assumed that the dust affects only the non-ionizing radiation in the galaxy. A similar assumption has been implemented in a number of other published articles that compute absolute escape fraction\cite{Steideletal2018}. Dust attenuation in the ionizing spectrum of galaxies remains a complex problem as it would depend on several factors of dust properties that are hard to constrain in observation. For example, dust fraction itself could be low in highly star-forming metal-poor galaxies \cite{Fisheretal2014} or there could be transparent holes in the ISM \cite{Reddyetal2016} through which ionizing radiation would escape without significant attenuation. 
Note that the original equation \cite{Leitetetal2013,Izotovetal2016a} uses f900 as a proxy for the ionizing flux. This paper uses f600 flux $\sim$ average flux density at $537 - 723$ \AA\, for the same (as in Eq.~\ref{eq:fesc600}).
The observed ratio of the flux densities at these wavelengths is obtained as $(f_{600}/f_{1500})^{obs} = 1.045$ and from the BPASS SED modelling, we have $(f_{600}/f_{1500})^{int} = 1.17$ for the age of the population, $4.5$~Myr.
The attenuation is obtained following Reddy extinction curve\cite{Reddyetal2016} at $\lambda =1500 \AA$ and is given by $A_{\lambda}(1500) = k_{\lambda}(1500) E(B-V) = 8.73\times E(B-V)$.
 {\it Then using $E(B-V) =0.13$, we obtain $f_{\rm esc} = 0.31$ (for $\tau_{IGM}=0$)  and $f_{\rm esc} = 0.42$ (for $\tau_{IGM}=0.29$). }

 The precise value is strongly dependent on the assumed intrinsic value of f600. The whole distribution of escape fractions for our best-guess stellar model is shown in Fig.~\ref{fig:EDFig4}(b). It peaks around 0.34; and 90\% of the sightlines have $0.31 < f_{\rm esc} < 0.42$. The right panel also shows $f_{\rm esc}$ distribution for other BPASS models with varying ages. We have derived escape fractions from three different methods and get consistent results, so the true escape fraction of rest-frame 600\AA ~LyC radiation from AUDFs01 is certainly $>20$\%, given the actual knowledge of stellar templates in extreme UV. 

\medskip

\noindent {\bf Data availability}\\
The HST data are available at {\url{https://3dhst.research.yale.edu/Data.php and\\ https://archive.stsci.edu/prepds/hlf}}. The VLT/ISAAC H and Ks band data are available at ESO Science Archive Facility (\url{http://archive.eso.org/scienceportal/home}). The Spitzer GOODS South data used in the analysis are available from {\url{https://irsa.ipac.caltech.edu/data/SPITZER/GOODS}}.\\
The SDSS data are available at the Sloan Digital Sky Survey (\url{https://www.sdss.org}). The MUSE spectroscopic data for AUDFs01 is available 
The other data that support the plots within this paper and other findings of this study are available from the corresponding author upon reasonable request.

\medskip

\noindent {\bf Code availability}\\
We have used standard data reduction tools in Python, IDL, IRAF, and the publicly available code SExtractor ({\url{https://www.astromatic.net/software/sextractor}}) for this study. For SED fitting and analysis, we have used publicly available code CIGALE 
({\url{https://cigale.lam.fr}}), EASY ({\url{http://www.astro.yale.edu/eazy/}}) and BPASS (\url{https://bpass.auckland.ac.nz/2.html}). The photoionization code  CLOUDY used in this paper is in public domain ({\url{https://trac.nublado.org/}}). The pipeline used to process the Level 1 AstroSat/UVIT data can be downloaded from {\url{http://astrosat-ssc.iucaa.in}}.

\medskip


\end{document}